\documentclass[12pt]{article}

\usepackage{scicite}
\usepackage{times}

\usepackage{aas_macros}
\usepackage{amssymb}
\usepackage{amsmath}
\usepackage{wasysym}
\usepackage{graphicx}
\usepackage{xspace}
\usepackage[usenames, dvipsnames]{color}
\usepackage[normalem]{ulem}
\usepackage{soul}
\usepackage{nicefrac}
\usepackage{makecell}
\usepackage{url}
\usepackage{xr-hyper}
\usepackage{hyperref}
\usepackage{footmisc}
\usepackage{pgffor}
\usepackage{subfig}
\usepackage[font=footnotesize]{caption}

\definecolor{dg}{rgb}{0.0, 0.6, 0.1}
\definecolor{brown}{rgb}{0.64705882, 0.16470588, 0.16470588}

\makeatletter

\def\myd{\mathrm{d}}
\def\myd{{\rm d}}
\def\dif{\@ifnextchar[{\@with}{\@without}}

\def\@with[#1]#2{
  \ensuremath{
    \mathchoice
    {\frac{\foreach \x in {#2}{\,\myd\x}}{\foreach \x in {#1}{\myd\x\,}}}%
    {{\foreach \x in {#2}{\,\myd\x}}/{\foreach \x in {#1}{\,\myd\x}}}%
    {{\foreach \x in {#2}{\,\myd\x}}/{\foreach \x in {#1}{\,\myd\x}}}%
    {{\foreach \x in {#2}{\,\myd\x}}/{\foreach \x in {#1}{\,\myd\x}}}
  }
}

\def\@without#1{
  \ensuremath{%
    \ifx\hfuzz#1\hfuzz
    \myd
    \else
    \foreach \x in {#1}{\,\myd\x}
    \fi
    }
}

\makeatother

\newcommand{\be}{\begin{equation}}
\newcommand{\ee}{\end{equation}}

\newcommand{\iso}{\ensuremath{^{\rm iso}}}
\newcommand{\intr}{\ensuremath{^{\rm int}}}
\newcommand{\obs}{\ensuremath{^{\rm obs}}}
\newcommand{\vhe}{\ensuremath{_{\textsc{vhe}}}}
\newcommand{\XRT}{\ensuremath{_{\textsc{xrt}}}}
\newcommand{\bat}{\ensuremath{_{\textsc{bat}}}}
\newcommand{\gbm}{\ensuremath{_{\textsc{gbm}}}}

\newcommand{\Fermi}{\emph{Fermi}\xspace}
\newcommand{\Swift}{\emph{Swift}\xspace}
\newcommand{\xrt}{\emph{Swift}-XRT\xspace}
\newcommand{\BAT}{\emph{Swift}-BAT\xspace}

\newcommand{\lat}{\emph{Fermi}-LAT\xspace}
\newcommand{\hess}{H.E.S.S.\xspace}
\newcommand{\grb}{GRB~190829A\xspace}
\newcommand{\lc}{light curve\xspace}
\newcommand{\lcs}{light curves\xspace}

\newcommand{\hr}{hrs\xspace}

\newcommand{\photonindex}{\gamma}
\newcommand{\tzero}{$T_0$}
\newcommand{\sct}[2]{{#1}$\times$10$^{#2}$}

\usepackage{booktabs,multirow}
\usepackage{floatrow}
\DeclareFloatFont{tiny}{\scriptsize}
\newenvironment{sciabstract}{%
\begin{quote} \bf}
{\end{quote}}

\title{Revealing X-ray and gamma ray temporal and spectral similarities in the GRB~190829A afterglow}
  
\author{H.E.S.S. Collaboration\footnote{Email: andrew.taylor@desy.de, Felix.Aharonian@mpi-hd.mpg.de, cromoli@mpi-hd.mpg.de, d.khangulyan@rikkyo.ac.jp, edna.ruiz@mpi-hd.mpg.de, fabian.schussler@cea.fr, sylvia.zhu@desy.de, contact.hess@hess-experiment.eu}
\footnote{H.E.S.S. Collaboration authors and affiliations are listed in the supplementary materials}}

\date{}

\begin{document}

\baselineskip24pt

\maketitle

\begin{sciabstract}

Gamma-Ray Bursts (GRBs), bright flashes of gamma rays from extragalactic sources followed by fading afterglow emission, are associated with stellar core collapse events. We report the detection of  very-high-energy (VHE) gamma rays from the afterglow of \grb, between 4 and 56~hrs after the trigger, using the High Energy Stereoscopic System. The low luminosity and redshift of \grb reduce both internal and external absorption, allowing determination of its intrinsic energy spectrum.
Between energies of 0.18 and 3.3~teraelectronvolts, this spectrum is described by a power law with photon index of $2.07\pm 0.09$, similar to the X-ray spectrum.
The X-ray and VHE gamma-ray \lcs also show similar decay profiles. 
These similar characteristics in the X-ray and gamma-ray bands challenge GRB afterglow emission scenarios.

\end{sciabstract}


The core collapse of a rapidly rotating massive star produces a supernova explosion, accompanied by a fast jet-like outflow propagating close to the speed of light. These outflows produce long GRBs, observed as prompt gamma-ray emission episodes lasting for seconds. This is followed by an afterglow of slowly fading emission, caused by the interaction of the relativistic ejecta with surrounding gas. The first radiative signature of GRB afterglows is non-thermal synchrotron emission from electrons accelerated at the forward shock of the relativistic outflow\cite{1992MNRAS.257P..29M}.  The same electrons up-scatter the synchrotron photons via the inverse Compton mechanism, producing a  synchrotron-self-Compton (SSC) emission component extending into the VHE ($>100$~GeV) regime.
This emission has been proposed as a second radiative signature of GRB afterglows
\cite{2001ApJ...548..787S,2001ApJ...559..110Z}. 
Two VHE GRB detections\cite{2019Natur.575..455M,2019Natur.575..459M,2019Natur.575..464A} have separately probed the early and late-time afterglow phases.

We observed the afterglow of \grb using the High Energy Stereoscopic System (\hess) on three consecutive nights, from 4.3 to 55.9~\hr after the GRB began. \grb was initially detected by the \Fermi Gamma-ray Burst Monitor (GBM), on 2019 August 29 at 19:55:53 universal time (UT) (\tzero)~\cite{GCN25551}. The Neil Gehrels \Swift Burst Alert Telescope (BAT) triggered on this burst 51~s later, and \Swift X-Ray Telescope (XRT) observations began 97.3~s after the BAT trigger~\cite{GCN25552}.

The \emph{Swift} Ultraviolet/Optical Telescope began observing the GRB afterglow at \tzero+158~s. The afterglow was also detected by ground-based telescopes in the optical, the near-infrared (NIR) and the radio bands, starting 1318~s after the GBM trigger and continuing for several days. The afterglow brightened in the optical-to-NIR bands starting approximately four days after \tzero ~due to the supernova associated with this GRB~\cite{GCN25623,GCN25651,GCN25664,GCN25677,GCN25682}. A host galaxy redshift of $z = 0.0785 \pm 0.0005$ was measured~\cite{2020arXiv200904021H}.

The first hour of the X-ray \lc has three flares/re-brightening events. The prompt emission is composed of a short peak at 0 to 4~s after \tzero ~and a second broader peak at 50 to 60~s after \tzero, both detected by the \emph{Swift}-BAT. A third re-brightening at 1000 to 3000~s after \tzero was detected by \emph{Swift}-XRT. Following this final re-brightening, the X-ray light curve followed a smooth power-law decay, typical of the afterglow onset~\cite{GCN25568}. Our \hess observations were performed during this power-law decay afterglow phase, beginning at \tzero $+10^{4}$~s.

The \hess observations were on three consecutive nights: the first night starting at \tzero+4.3~\hr for a total of 3.6~\hr, the second night at  \tzero+27.2~\hr for 4.7~\hr, and the third night at \tzero+51.2~\hr for 4.7~\hr \cite{SOM}. Our analysis detected a VHE gamma-ray signal on each of the three nights, spatially coincident with the GRB position (Fig.~\ref{fig:S0}), with statistical significances of 21.7\,$\sigma$, 5.5\,$\sigma$ and 2.4\,$\sigma$ \cite{SOM}.

\begin{figure}[ht!]
  \includegraphics[width=1.0\textwidth]{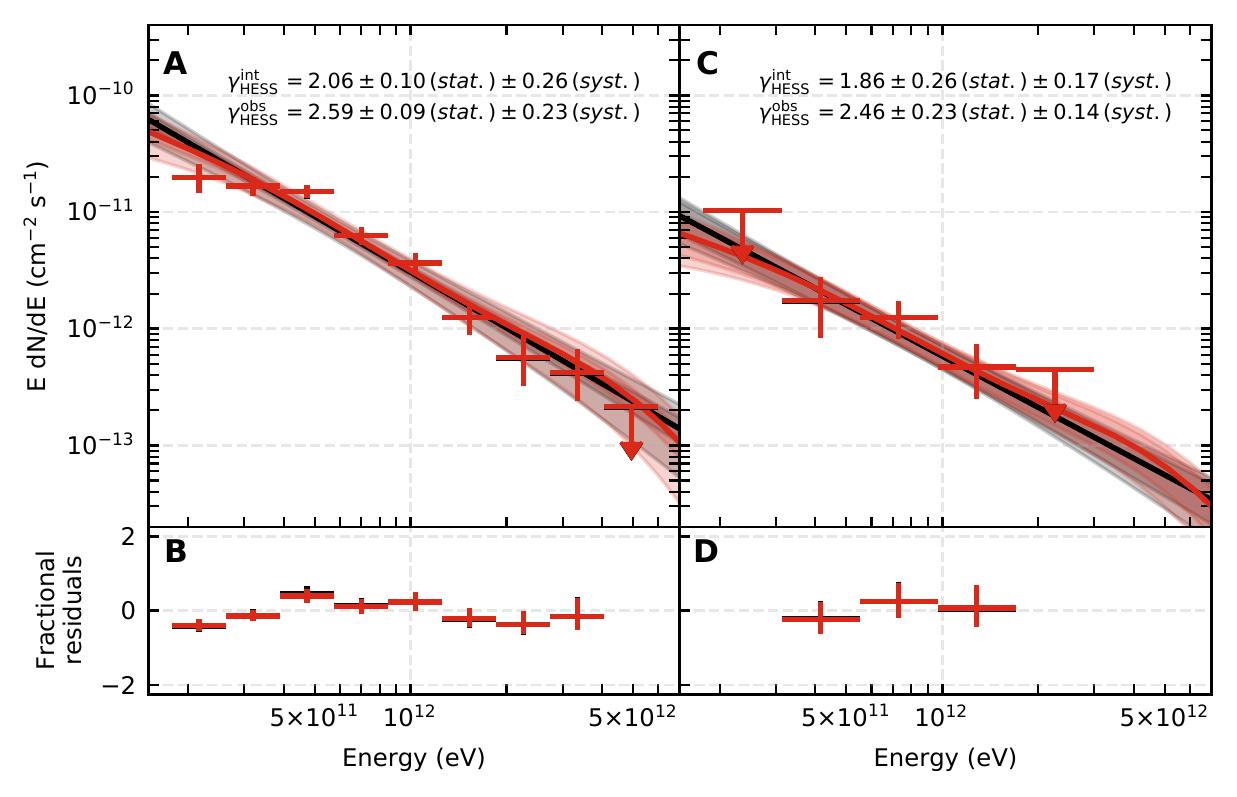}
      \caption{{\bf \hess VHE spectra of \grb on the first and second nights}. (A) the power-law (black) and EBL attenuated power-law (red) models (lines) fitted to the observational data (red crosses) with shaded regions indicating the 1\,$\sigma$ statistical and systematic uncertainty on the first night. (C) the same as panel A but for the second night of observations. (B and D) show the fractional residuals between the data and the power-law (black) and EBL attenuated power-law (red) models, defined as (data-model)/model. Error bars in all panels are 1\,$\sigma$ statistical uncertainty, and upper limits are the 95\% confidence level.}
  \label{fig:spectrum}
\end{figure}

We performed a spectral analysis of the first two nights; the signal on the third night was too weak to determine the spectrum. 
We fitted a power-law model to the photon spectrum of the form $dN/dE=N_{0}(E/E_{0})^{-\photonindex\obs\vhe}$, where $N_{0}$ is the spectrum normalisation at photon energy $E_0$, and $\photonindex\obs\vhe$ is the spectral index. We find $\photonindex\obs\vhe =2.59~\pm~0.09$\,(stat.)\,$\pm~0.23$\,(syst.) (0.18\,TeV to 3.3\,TeV) on the first night and $2.46~\pm~0.22$\,(stat.)\,$\pm~0.14$\,(syst.) (0.18\,TeV to 1.4\,TeV) on the second night (Fig.~\ref{fig:spectrum}). 

\begin{figure}[ht]
    \centering
    \includegraphics[width=0.5\textwidth]{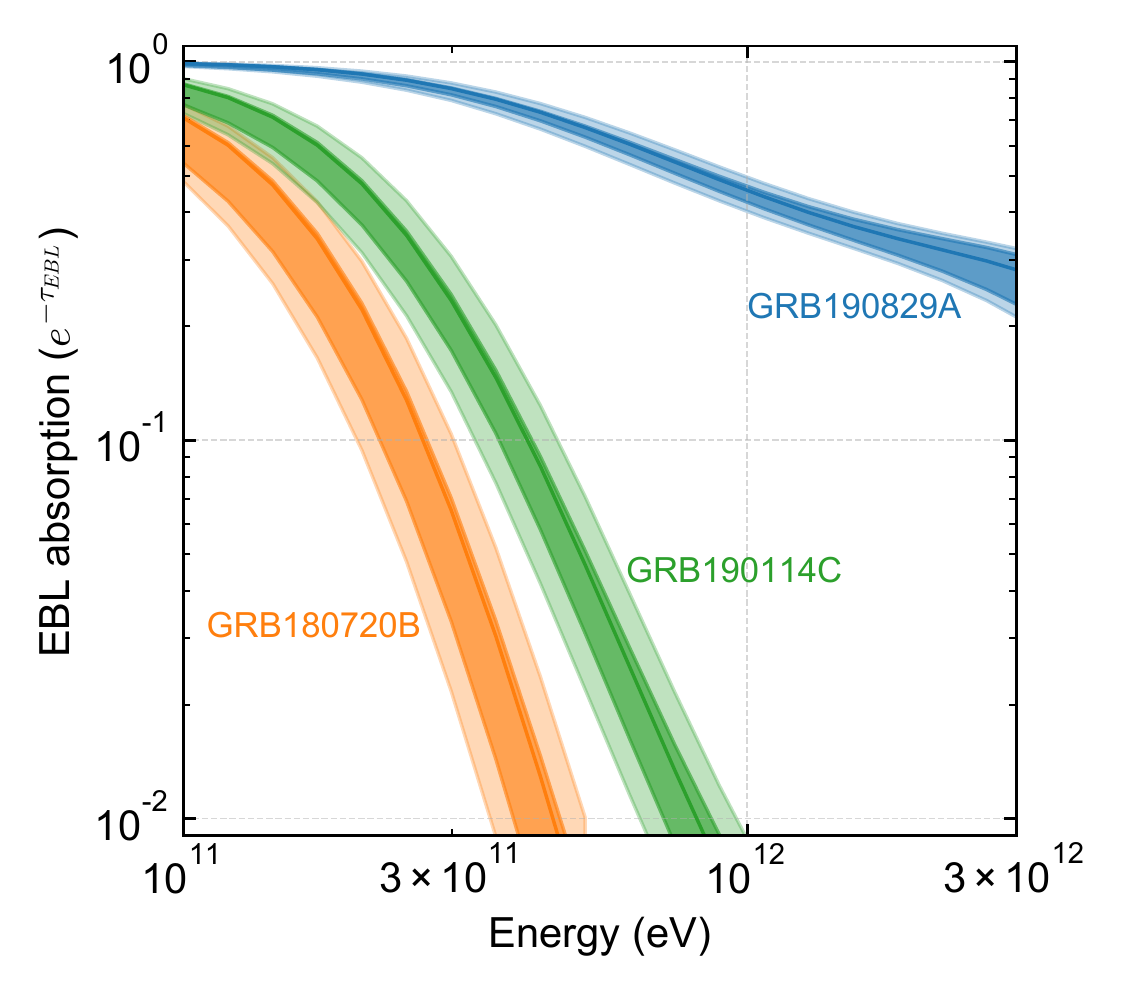}
    \caption{ {\bf Extragalactic background light absorption for three GRBs detected in the VHE band}. Models of EBL absorption in the energy range 0.1 to 3 TeV are shown for \grb ($z=0.0785$, blue \cite{2020arXiv200904021H}), for GRB~190114C ($z=0.4245$ \cite{2019Natur.575..455M}, green), and for GRB~180720B ($z=0.653$ \cite{2019Natur.575..464A}, orange). The solid lines indicate the EBL absorption from our preferred model~\cite{Franceschini2008}, the darker shaded areas indicate the spread when adopting EBL models~\cite{Franceschini2017,Finke2010,Dominguez2011,Gilmore 2012}, and lighter shaded areas indicate the additional uncertainty from a shift of 10\% in the energy scale.}
    \label{fig:absorption_ebl_int}
\end{figure}

Gamma rays from distant astronomical sources are attenuated by photon interactions with radiation fields both within the source (``internal'' absorption), and with the extragalactic background light (EBL)\cite{Franceschini2008}. Whilst extragalactic absorption depends on redshift $z$, internal absorption depends on the source compactness (the ratio of the intrinsic source luminosity to its radius). The modest luminosity of \grb \cite{2019GCN.25575....1L} and late observation epochs reduce the internal absorption, so that the attenuation on the EBL is expected to dominate in the \hess observation period\cite{SOM}.  
The gamma-ray attenuation by the EBL increases with source distance; the Universe is optically thick to multi TeV gamma rays beyond a redshift of $z\approx 0.1$  \cite{Franceschini2008}. Nearby GRBs experience less attenuation, requiring a smaller correction due to EBL absorption and allowing the intrinsic spectrum to be determined. Fig.~\ref{fig:absorption_ebl_int} shows the EBL absorption level for \grb and for the other two GRBs detected at VHEs \cite{2019Natur.575..464A,2019Natur.575..455M}.

To characterise the intrinsic GRB spectrum (corrected for EBL absorption), we fitted an attenuated power-law model of the form $\dif[E]{N} = N_{0} (E/E_{0})^{-\photonindex\intr\vhe} e^{-\tau(E,z)}$ to the data. The exponential term corresponds to the absorption of photons through their interaction with the EBL, and  $\tau$ is the energy-dependent optical depth for a source at  redshift $z$ \cite{SOM}. 

For the first two observation nights, we determined VHE intrinsic photon indices of:  $\photonindex\intr\vhe =  2.06\pm 0.10$\,(stat.)\,$\pm 0.26$\,(syst.) (1st night), $1.86\pm 0.26$\,(stat.)\,$\pm 0.17$\,(syst.) (2nd night). These values indicate that, within the uncertainties, the spectral shape remains unchanged, so we performed a joint spectral analysis \cite{SOM}.  Combining the three nights of observation data, the photon index found is  $\photonindex\vhe\intr$=$2.07\pm 0.09$\,(stat.)\,$\pm 0.23$\,(syst.), in the energy range (0.18 - 3.3\,TeV). These per night VHE photon indices are consistent, within the statistical uncertainties,  with the photon indices of the X-ray emission ($\photonindex\XRT$) we measured from the \xrt data taken during the same observational periods: $\photonindex\XRT=2.03\pm 0.06$ (1st night); $\photonindex\XRT=2.04\pm 0.10$ (2nd night) \cite{SOM}.\newline

\begin{figure}[ht!]
    \includegraphics[width=1.0\textwidth]{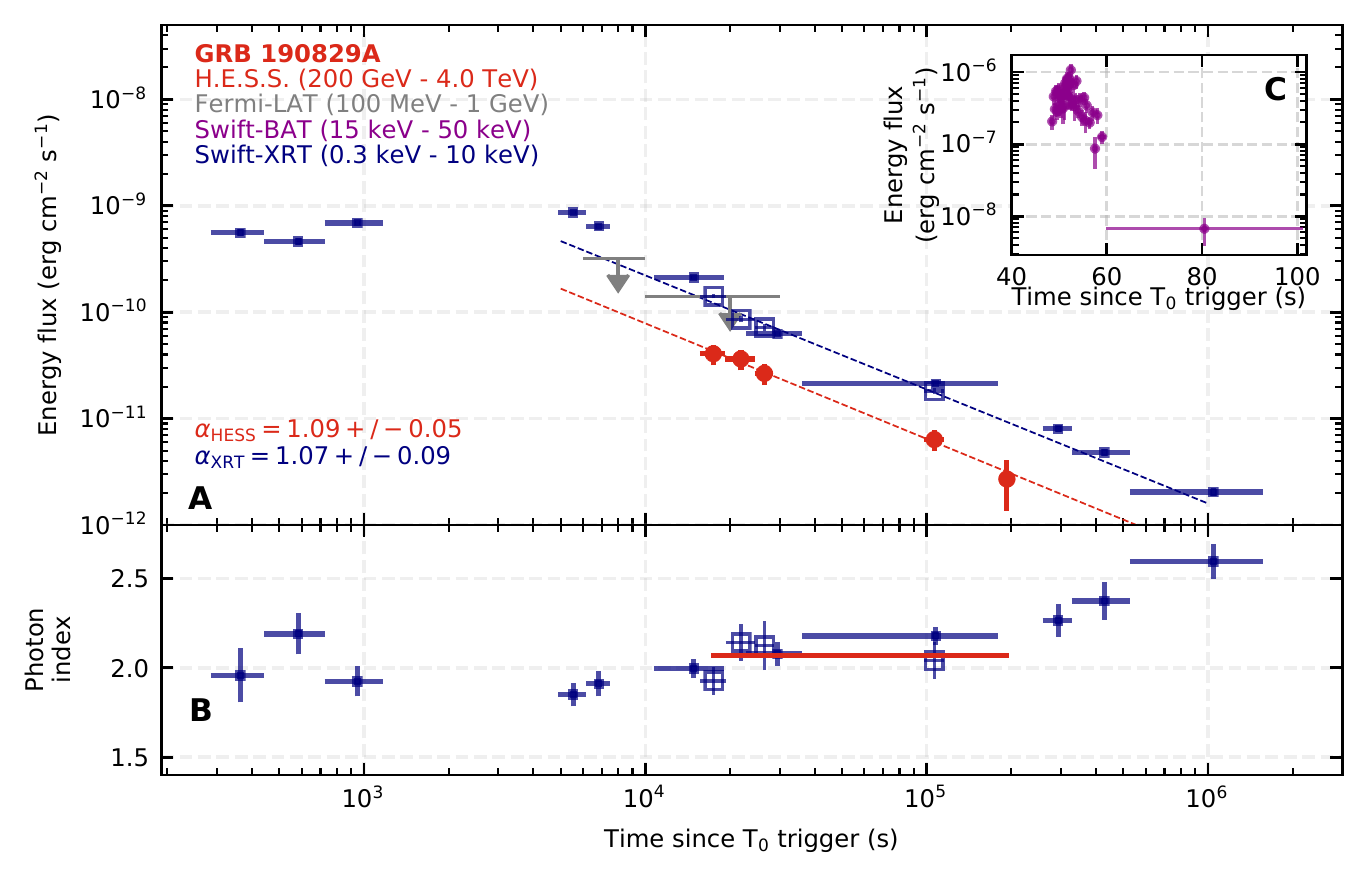}
    \caption{{\bf Logarithmic X-ray and gamma-ray multi-wavelength energy flux \lcs of the \grb afterglow}. (A) the temporal evolution of the energy flux detected in X-rays by \xrt (blue closed squares), upper limits on MeV gamma rays by \lat (grey arrows) and VHE gamma rays by \hess (red circles). The XRT temporal decay index ($\alpha\XRT$) was obtained by fitting a model to only the XRT data that were  simultaneous with the \hess observations (open squares). (B) the corresponding intrinsic photon indices. The \hess intrinsic spectral index, indicated by the continuous red line, is assumed to be constant at the mean value of $2.07~\pm~0.09$ determined from nights 1 to 3. (C) the energy flux evolution of the prompt emission observed by \BAT, obtained from the \emph{Swift} Burst Analyser \cite{2010A&A...519A.102E}. All error bars correspond to 1\,$\sigma$ uncertainty, and the \emph{Fermi-LAT} upper limits are at the 95\% confidence level.}
    \label{fig:mwl_lightcurve}
\end{figure}

A \lc of the \hess observations was extracted in the 0.2 to 4.0~TeV energy range for the entire temporal coverage up to \tzero+56~\hr. We split the first observation night into three sub-intervals (Fig.~\ref{fig:mwl_lightcurve}). The gamma-ray energy flux, $F_\mathrm{VHE}$, depends on how much time $t$ has passed after $T_0$, and the time evolution is characterised by a power-law model  $F_\mathrm{VHE}(t)\propto t^{-\alpha_{\vhe}}$, with $\alpha\vhe=1.09\pm 0.05$ \cite{SOM}. This VHE gamma-ray flux behaviour is similar to the X-ray \lc derived for the same time period. The flux measured by \xrt can also be described as a power law, with index $\alpha_\mathrm{XRT}$; using the \xrt data in the energy interval 0.3--10~keV, we find that $\alpha\XRT=1.07\pm 0.09$.

The photon index of the \grb afterglow in the X-ray band is typical for GRB afterglows, which have a mean value of $\bar{\photonindex}\XRT\approx 2$ \cite{FermiLAT:2018ksx}. However, the afterglow decayed more slowly than  is typical over this time interval. The average decay index for all GRB afterglows measured by \xrt up to 56~\hr post \tzero ~time period (observer frame) is $\bar{\alpha}\XRT \sim 1.4$, based on the best-fitting broken power-law \lc models, with less than 30\% of the XRT-detected GRBs having smaller temporal decay indices than \grb, during the same time period \cite{SOM}.

Fig.~\ref{fig:mwl_lightcurve} shows the X-ray and gamma-ray afterglow temporal evolutions between 20~s and 11~days after \tzero. The afterglow phase, $t \geq 10^3$~s, shows the \hess and \xrt detected fluxes and upper limits on the gamma-ray flux at 0.1 to 1\,GeV, derived from the \lat observations~\cite{GCN25574}. During this afterglow period, the temporal profiles in the X-ray band show no short-timescale variability. The VHE \lc closely matches the X-ray \lc \cite{SOM}.

During the prompt phase ($t<100~{\rm s}$), the total isotropic energy in the GBM energy band (10 -- 1000~keV) was $E\gbm\iso \approx 2\times10^{50}$~erg with an uncertainty at the 1\% level \cite{2019GCN.25575....1L}, and in the \BAT energy range (15 -- 150\,keV) was $E\bat\iso\approx 1\times 10^{50}$~erg with an uncertainty at the 10\% level ($t<60~{\rm s}$). With the energy flux in the afterglow decaying at approximately $t^{-1}$, the energy output per logarithmic time interval is approximately constant. Consequently, no single logarithmic temporal bin dominates the total fluence of the outburst (Fig.~\ref{fig:fluence_evol}). 
The isotropic energy output measured by \xrt during the afterglow phase was $E\XRT\iso\approx 5\times 10^{50}$~erg ($t<10^{6}~{\rm s}$) with an uncertainty at the 10\% level \cite{SOM}. This is larger than during the prompt phase, a feature that has rarely been observed \cite{2014MNRAS.444..250E}.

We investigated whether the  VHE light curve and spectrum of \grb are compatible with the standard GRB afterglow emission model. This model assumes particle acceleration at the forward shock \cite{SOM}, where the outflow propagates outwards through the circumburst material \cite{1976PhFl...19.1130B}. As a result, a non-negligible fraction of the shock energy is transferred into magnetic field enhancement and particle acceleration, leading to the production of broadband non-thermal emission.

Although both hadrons and leptons are accelerated, typical values of the circumburst density at the forward shock suggest that the time for hadrons to cool is substantially longer than the dynamical shock time, resulting in rather low radiative efficiencies. Unlike hadrons, VHE electrons promptly lose their energy through synchrotron and inverse Compton radiation.

The observed energy flux decays as approximately \(t^{-1}\) in both X-ray and gamma-ray bands.
Decay of this form suggests that the shocked plasma magnetisation level, the fraction of energy transferred to non-thermal electrons, and radiative efficiency of the forward shock effectively remain constant during the afterglow \cite{SOM}. 

We performed multi-wavelength modelling of the \xrt, \lat and \hess data. We averaged these spectral results on a per night basis since this is comparable to the evolution time scale of the outflow Lorentz factor, dictated by GRB kinematics \cite{1976PhFl...19.1130B}. We find that the emission region has a Lorentz factor of $\Gamma = 4.7$ and $\Gamma = 2.6$ for the first and second nights, respectively. 

For Lorentz factors $<10$, expected during the \hess observation window\cite{SOM}, the accelerated electrons producing the VHE emission experience recoil when up-scattering the synchrotron photons, referred to as the onset of the Klein-Nishina regime\cite{SOM}, resulting in steepening of the inverse Compton spectrum. This steepening makes it challenging for SSC models to reproduce simultaneously the observed X-ray and VHE spectra. We therefore introduced an alternative {leptonic} scenario with no limitation placed on the electron maximum energy (which would otherwise be set by energy losses). This assumption implies that the synchrotron spectrum can extend up into the VHE regime. 

\begin{figure}[ht!]
  \includegraphics[width=\textwidth]{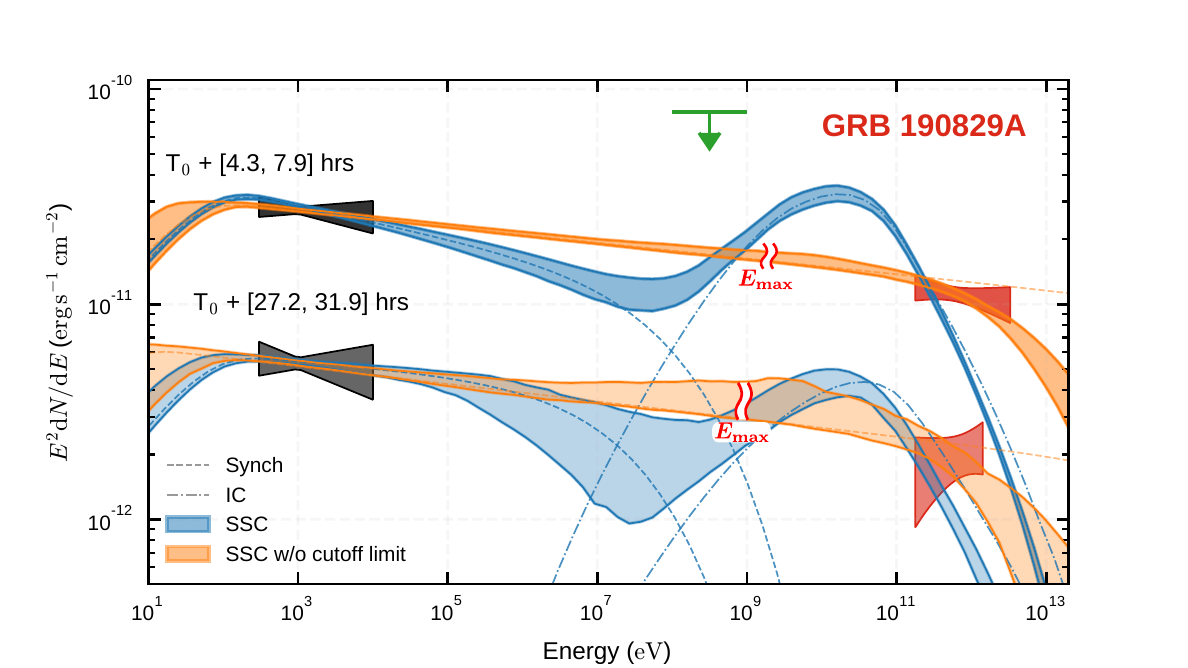}
  \caption{{\bf Theoretical multi-wavelength models of the first and second night data}. The black region shows the spectrum and uncertainty of the \xrt data, the green arrow upper limit is from \lat (available only for the first night \cite{GCN25574}), and the red region is the \hess intrinsic spectrum and its uncertainty (statistical only). The shaded areas represent the 68\% confidence intervals determined from the posterior probability distribution of the MCMC parameter fitting for the standard SSC model (light blue) and for the synchrotron-dominated model (orange); the latter model does not impose a synchrotron cut-off energy (shown by \(E_{\rm max}\)). The synchrotron components of the two SSC models are indicated by dashed lines, while the dash-dotted lines show the inverse Compton components. These lines show the emission level when neglecting the internal gamma-gamma absorption. The upper curves are for the first night and the lower curves the second night. The best-fitting parameters are listed in Tables~\ref{tab:par_fullprior}-\ref{tab:par_nocutoff}\cite{SOM}.}
  \label{fig:MWL_fit_sync_limit}
\end{figure}

To further investigate the emission origin, we searched for a theoretical instantaneous electron distribution such that the corresponding synchrotron and SSC emission can explain consistently both the X-ray and gamma-ray spectra. We performed a Markov Chain Monte Carlo exploration of the five-dimensional (the magnetic field strength, and four parameters describing the broken power-law electron distribution) parameter space \cite{SOM}, with the results shown in Fig.~\ref{fig:MWL_fit_sync_limit}. Additionally, we investigated whether including the optical data\cite{2020arXiv200904021H} affects these results, finding that they remain unchanged\cite{SOM}.

The standard model in which the electron maximum energy is set by the energy loss limit  predicts a soft spectral index for the VHE emission. This is due to the combination of the accelerated electrons having a steep distribution (power-law index $\beta_2 \approx 3$) and the fact that in the VHE range the photons are produced via inverse Compton scattering in the Klein-Nishina regime. Internal photon-photon absorption within the source makes the spectrum steeper. Such a spectrum is inconsistent with our observations. 

For the alternative model with no limit placed on the maximum electron energy, the theoretical spectrum is dominated by a single synchrotron component covering a broad energy range from X-rays to VHE gamma-rays (Fig.~\ref{fig:MWL_fit_sync_limit}).  The SSC component in this case is 3 orders of magnitude weaker than the synchrotron component. In the VHE range covered by the \hess observations, internal photon-photon absorption is non-negligible. A single synchrotron component provides a significantly better ($>5$\,$\sigma$) fit to the multiwavelength data. 

However, if particle acceleration and emission occurs in a region where ideal magnetohydrodynamic conditions are satisfied, the synchrotron component should not extend beyond \(E_{\rm max}\approx 200 D \rm\,MeV\) (here the Doppler factor is \(D\approx2\Gamma\) for \(\Gamma\gg1\). Fig.~\ref{fig:MWL_fit_sync_limit} shows that the synchrotron component would need to extend more than three orders of magnitude beyond the synchrotron limiting energy. This would require an unknown high-efficiency process to accelerate multi PeV electrons in the magnetic fields (expected to be a few Gauss in strength) or a conventional acceleration mechanism in a medium with a large difference in the magnetic field strengths of the acceleration and radiation zones \cite{2012MNRAS.427L..40K}.

The spectral steepening predicted in the VHE range means we cannot reproduce the observations with a simple one-zone SSC models (Fig. \ref{fig:MWL_fit_sync_limit}). We discuss two ways to improve the agreement between data and the models: (i) higher bulk Lorentz factor; and (ii) non-powerlaw distributions of emitting particles.  

The discrepancy between models and data could be removed if the Doppler factor at 4~\hr after \tzero ~is increased to $\sim 10^{2}$. This assumption reduces the level of electron recoil in the inverse Compton scattering process, bringing the intrinsic VHE photon index closer to that measured in X-rays; and the internal photon-photon absorption of the VHE emission is reduced. However, the required value of the bulk Lorentz factor is inconsistent with predictions of the standard hydrodynamic model. The weak dependence of the Lorentz factor on the explosion energy and surrounding gas density make $\sim 10^2$ values difficult to physically produce \cite{SOM}. If \grb was an off-axis explosion, with a large line-of-sight angle relative to the jetted outflow direction, the bulk Lorentz factor could have been underestimated. However, even in this case the bulk Lorentz factor on the second night would have been smaller, which is inconsistent with the unchanged observed spectrum \cite{SOM}.

If the accelerated electrons were to have an additional hard energy distribution at high energies, then the synchrotron and inverse Compton components can potentially explain the X-ray and gamma-ray spectra recorded from \grb. However, this would require extreme assumptions regarding the properties of the circumburst medium.  If inverse Compton losses are dominant, the electrons which are cooled by inverse Compton losses follow an energy distribution that is harder than the injected distribution \cite{SOM}. However, this requires that the spectral energy distribution is strongly dominated by the inverse Compton component, which is inconsistent with the observed TeV flux and  \lat upper limit (Fig.~\ref{fig:MWL_fit_sync_limit}). 

In summary, we measured a hard intrinsic spectral index, $\photonindex\intr\vhe\approx 2$, over more than an order of magnitude in energy in the VHE band in the late time afterglow, \tzero +4~\hr to 56~\hr. The gamma-ray energy flux and index appears consistent with an extrapolation of the synchrotron spectrum observed in the X-ray band. This simple spectral behaviour proves difficult to describe with simple one zone emission models.

\newpage




\newpage

\nocite{HESSCrab}
\nocite{HESSIIcrab}
\nocite{HESSGRBs}
\nocite{GCN25555}
\nocite{bgmodeling}
\nocite{ImPACT}
\nocite{TMVAHESS}
\nocite{MohrmannOpenSource}
\nocite{gammapy:2019}
\nocite{gammapy:2017}
\nocite{deNaurois:2009ud}
\nocite{Gillessen:2004tc}
\nocite{refId0}
\nocite{HESSEBL}
\nocite{2007A&A...469..379E}
\nocite{Evans09}
\nocite{Arnaud96}
\nocite{Wilms00}
\nocite{Verner96}
\nocite{Willingale13}
\nocite{2019ApJ...880L..27D}
\nocite{Landau6}
\nocite{Sari1998}
\nocite{naima}
\nocite{2020MNRAS.496.3326R}
\nocite{2020arXiv200311252F}
\nocite{2020ApJ...898...42C}
\nocite{1983MNRAS.205..593G}
\nocite{2009ApJ...703..675N}
\nocite{1985Ap.....23..650A}
\nocite{2014ApJ...783..100K}
\nocite{2005AIPC..745..359K}
\nocite{1996MNRAS.278..525A}

\bibliographystyle{science}
\bibliography{main}
\section*{Acknowledgements}

\noindent {\bf Acknowledgements} 
We thank C.~Arcaro, N.~Zywucka, H.~Ashkar for discussions of the MWL modelling, and L.~Mohrmann for guidance on the \textsc{gammapy} analysis. 

\noindent {\bf Funding}
Supported by the Max Planck Society (E.~R.~V, C.~R., and F.~A.), the Dublin Institute for Advanced Studies (F.A.), the Helmholtz Association (A.~M.~T., S.~J.~Z.), the Japanese Society for the Promotion of Science (D.~K.), and the Commissariat \`a l'\'energie atomique et aux \'energies alternatives (F.~S.). The H.E.S.S. collaboration gratefully acknowledges the support of the Namibian authorities and of the University of Namibia in facilitating the construction and operation of H.E.S.S., as well as the support by the German Ministry for Education and Research (BMBF), the Max Planck Society, the German Research Foundation (DFG), the Helmholtz Association, the Alexander von Humboldt Foundation, the French Ministry of Higher Education, Research and Innovation, the Centre National de la Recherche Scientifique (CNRS/IN2P3 and CNRS/INSU), the Commissariat à l'énergie atomique et aux énergies alternatives (CEA), the U.K. Science and Technology Facilities Council (STFC), the Knut and Alice Wallenberg Foundation, the National Science Centre, Poland grant no. 2016/22/M/ST9/00382, the South African Department of Science and Technology and National Research Foundation, the University of Namibia, the National Commission on Research, Science \& Technology of Namibia (NCRST), the Austrian Federal Ministry of Education, Science and Research and the Austrian Science Fund (FWF), the Australian Research Council (ARC), the Japan Society for the Promotion of Science and by the University of Amsterdam. We appreciate the excellent work of the technical support staff in Berlin, Zeuthen, Heidelberg, Palaiseau, Paris, Saclay, Tübingen and in Namibia in the construction and operation of the equipment. This work benefited from services provided by the H.E.S.S. Virtual Organisation, supported by the national resource providers of the EGI Federation.

\noindent {\bf Author Contributions:} E.~L.~Ruiz-Velasco carried out the main H.E.S.S. data analysis, and F.~Schussler provided the cross-check. S.~Zhu led the X-ray analysis and discussion. C.~Romoli, D.~Khangulyan, F.~Aharonian, and A.~M.~Taylor performed modelling and interpretation. The manuscript was prepared by E.~L.~Ruiz-Velasco (supervised by J.~Hinton), S.~J.~Zhu, C.~Romoli, D.~Khangulyan, F.~Aharonian, and A.~M.~Taylor. S.~Wagner is the collaboration spokesperson. 
Other \hess collaboration authors contributed to the design, construction and operation of \hess, the development and maintenance of data handling, data reduction and data analysis software. All authors meet the journal's authorship criteria and have reviewed, discussed, and commented on the results and the manuscript.

\noindent {\bf Competing Interests:} The authors declare that they have no competing interests.

\noindent {\bf Data and Materials Availability:} The H.E.S.S. data are available at: \url{https://www.mpi-hd.mpg.de/hfm/HESS/pages/publications/auxiliary/GRB190829A/Auxiliary_Information_GRB190829A.html}. This includes the points of the spectral energy distributions (cf. Fig. 1), the light-curve (cf. Fig. 2), the significance skymaps (cf. Fig. S1), as well as the code used to derive the presented MCMC based modelling. The Swift-XRT data is available at \url{https://www.swift.ac.uk/xrt_spectra/00922968/} and the files to reproduce our temporal analysis are available at the same URL as the H.E.S.S. data.


\textbf{Supplementary Materials: Authors and affiliations, Materials and Methods, Figure S1-S9, Tables S1-S7, References (29-61)}

\begin{appendix}

\subsection*{H.E.S.S. Collaboration authors and affiliations}
H.~Abdalla$^{1}$,
F.~Aharonian$^{2,3,4}$,
F.~Ait~Benkhali$^{3}$,
E.O.~Ang\"uner$^{5}$,
C.~Arcaro$^{6}$,
C.~Armand$^{7}$,
T.~Armstrong$^{8}$,
H.~Ashkar$^{9}$,
M.~Backes$^{1,6}$
% ...
\\\newline
1. University of Namibia, Department of Physics, Windhoek 10005, Namibia\\
2. Dublin Institute for Advanced Studies, Dublin 2, Ireland\\
3. Max-Planck-Institut f\"ur Kernphysik, D 69029 Heidelberg, Germany\\
4. High Energy Astrophysics Laboratory, Russian-Armenian University (RAU), Yerevan 0051, Armenia\\
5. Aix Marseille Universit\'e, Centre national de la recherche scientifique (CNRS)/Institut National de Physique Nucléaire et Physique des Particules (IN2P3), Centre de Physique des Particules de Marseille (CPPM), Marseille, France\\
6. Centre for Space Research, North-West University, Potchefstroom 2520, South Africa\\
7. Laboratoire d'Annecy de Physique des Particules (LAPP), Univ. Grenoble Alpes, Univ. Savoie Mont Blanc, CNRS, 74000 Annecy, France\\
8. University of Oxford, Department of Physics, Denys Wilkinson Building, Oxford OX1 3RH, UK\\
9. Institute for Research on the Fundamental Laws of the Universe (IRFU), Commissariat à l'énergie atomique (CEA), Universit\'e Paris-Saclay, F-91191 Gif-sur-Yvette, France\\
10. Instytut Fizyki J\c{a}drowej Polskiej Akademii Nauk (PAN), 31-342 Krak{\'o}w, Poland\\
11. Deutsches Elektronen-Synchrotron (DESY), D-15738 Zeuthen, Germany\\
12. Obserwatorium Astronomiczne, Uniwersytet Jagiello{\'n}ski, 30-244 Krak{\'o}w, Poland\\
13. Department of Physics and Electrical Engineering, Linnaeus University,  351 95 V\"axj\"o, Sweden\\
14. Institut f\"ur Astronomie und Astrophysik, Universit\"at T\"ubingen, D 72076 T\"ubingen, Germany\\
15. Dipartimento Interateneo di Fisica, Politecnico di Bari, 70125 Bari, Italy\\
16. Istituto Nazionale di Fisica Nucleare, Sezione di Bari, 70125 Bari, Italy\\
17. Laboratoire Univers et Th\'{e}ories, Observatoire de Paris, Universit\'{e} PSL, CNRS, Universit\'{e} de Paris, 92190 Meudon, France\\
18. Sorbonne Universit\'e, Universit\'e Paris Diderot, Sorbonne Paris Cit\'e, CNRS/IN2P3, Laboratoire de Physique Nucl\'eaire et de Hautes Energies (LPNHE), F-75252 Paris, France\\
19. Gravitation and Astroparticle Physics at the University of Amsterdam (GRAPPA), Anton Pannekoek Institute for Astronomy, University of Amsterdam, 1098 XH Amsterdam, The Netherlands\\
20. Friedrich-Alexander-Universit\"at Erlangen-N\"urnberg, Erlangen Centre for Astroparticle Physics, D 91058 Erlangen, Germany\\
21. Astronomical Observatory, The University of Warsaw, 00-478 Warsaw, Poland\\
22. School of Physics, University of the Witwatersrand, Braamfontein, Johannesburg, 2050 South Africa\\
23. Universit\'{e} de Paris, CNRS, Astroparticule et Cosmologie, F-75013 Paris, France\\
24. Universit\"at Hamburg, Institut f\"ur Experimentalphysik, D 22761 Hamburg, Germany\\
25. School of Physics and Astronomy, The University of Leicester, Leicester, LE1 7RH, United Kingdom\\
26. Nicolaus Copernicus Astronomical Center, Polish Academy of Sciences, 00-716 Warsaw, Poland\\
27. Institut f\"ur Physik und Astronomie, Universit\"at Potsdam, D 14476 Potsdam, Germany\\
28. School of Physical Sciences, University of Adelaide, Adelaide 5005, Australia\\
29. Laboratoire Leprince-Ringuet, CNRS, Institut Polytechnique de Paris, F-91128 Palaiseau, France\\
30. Laboratoire Univers et Particules de Montpellier, Universit\'e Montpellier, CNRS/IN2P3, F-34095 Montpellier Cedex 5, France\\
31. Universit\'e Bordeaux, CNRS/IN2P3, Centre d'\'Etudes Nucl\'eaires de Bordeaux Gradignan, 33175 Gradignan, France\\
32. Institut f\"ur Astro- und Teilchenphysik, Leopold-Franzens-Universit\"at Innsbruck, A-6020 Innsbruck, Austria\\
33. Landessternwarte, Universit\"at Heidelberg, K\"onigstuhl, D 69117 Heidelberg, Germany\\
34. Institute of Astronomy, Faculty of Physics, Astronomy and Informatics, Nicolaus Copernicus University, 87-100 Torun, Poland\\
35. Department of Physics, Rikkyo University, Toshima-ku, Tokyo 171-8501, Japan\\
36. Institut f\"ur Physik, Humboldt-Universit\"at zu Berlin, D 12489 Berlin, Germany\\
37. Department of Physics, University of the Free State, Bloemfontein 9300, South Africa\\
38. Department of Physics, The University of Tokyo, Bunkyo-ku, Tokyo 113-0033, Japan\\
39. Yerevan Physics Institute, 375036 Yerevan, Armenia\\
40. Kavli Institute for the Physics and Mathematics of the Universe (World Premier International Research Center Initiative (WPI)), The University of Tokyo Institutes for Advanced Study (UTIAS), The University of Tokyo, Kashiwa, Chiba, 277-8583, Japan\\
41. School of Physics and Astronomy, Sun Yat Sen University, Guangzhou 510275, People’s Republic of China\\

\renewcommand{\thetable}{S\arabic{table}}
\renewcommand{\thefigure}{S\arabic{figure}}
\renewcommand{\theequation}{S\arabic{equation}}


\setcounter{figure}{0}

\section*{Materials and methods}

\subsection*{H.E.S.S. observations and data analysis}

\begin{figure}[h]
    \centering
    \includegraphics[width = 1.0\linewidth]{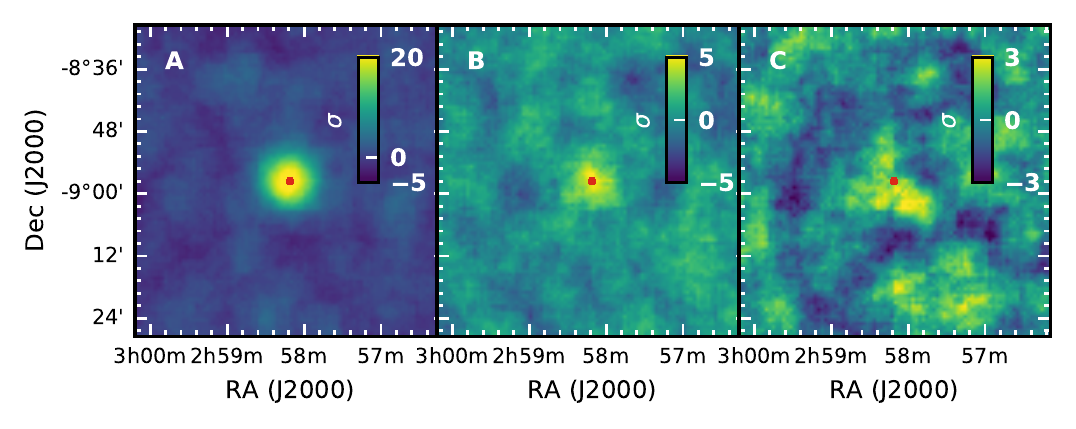}
    \caption{\textbf{Significance skymap of the H.E.S.S. observations.} Skymaps in equatorial coordinates of the H.E.S.S. observations in the GRB region for night 1, 2 and 3 in panel A, B and C respectively. The red cross shows the fitted point-source position corresponding to right ascension 02$\rm ^h$58$\rm ^{min}$11.1$\rm ^s$ and declination $-$08$^\circ$58$'$4.8$''$ (J2000).}
    \label{fig:S0}
\end{figure}

\hess is an array of five Imaging Atmospheric Cherenkov Telescopes located in the Khomas Highland of Namibia, designed to detect gamma-ray sources in the tens of GeV to tens of TeV energy range \cite{HESSCrab,HESSIIcrab}. \hess observations of GRBs are triggered by external alerts and follow-up is performed based on specific trigger criteria, on a case-by-case basis accounting for the redshift, X-ray flux, and delay of the GRB follow-up\cite{HESSGRBs}. \grb was observed with the \hess array for a total of 12.98~hrs distributed over three nights, starting at \tzero+4.3~hrs and finishing at \tzero+55.9~hrs. The observations were performed using stereoscopic observations with CT1-4 (the four medium-size telescopes of \hess); due to a maintenance campaign, the large telescope (CT5) was only functioning intermittently. The observations were taken in "wobble mode" with an offset of 0.5 degrees from the position of the GRB (right ascension 02$\rm ^h$58$\rm ^{min}$10.6$\rm ^s$, declination $-$08$^\circ$57$'$29.8$''$ (J2000 equinox))~\cite{GCN25555}. The reflected-region method was used for the determination of the background regions and extraction of spectral results and ring-region for the extraction of skymaps shown in Fig.\ref{fig:S0}.~\cite{bgmodeling}. 
The reconstruction of the shower properties
was done using the \textsc{ImPACT}~\cite{ImPACT} maximum likelihood-based technique and background events were rejected with a multivariate analysis scheme\cite{TMVAHESS}. The spectral fitting procedure was performed with a likelihood model fitting, using the \textsc{gammapy} software package\cite{MohrmannOpenSource,gammapy:2019,gammapy:2017} version 0.17. The results were cross-checked with an independent calibration and analysis chain~\cite{deNaurois:2009ud}. 

\subsubsection*{Source position}
The gamma-ray excess position detected by \hess was fit with a point-like source model convolved with the point-spread function of the analysis (0.07$^\circ$, \cite{ImPACT}) in the significance skymap obtained from the three observations night together. The fit results in a source position centred at right ascension  02$\rm ^h$58$\rm ^{min}$11.1$\rm ^s$ and declination $-$08$^\circ$58$'$4.8$''$ (J2000 equinox) (Fig.\ref{fig:S0}) with a statistical uncertainty of 11.4$''$, which is 35.9$''$ away from the GRB position measured in the optical band~\cite{GCN25555}. H.E.S.S has a systematic pointing uncertainty of about 20$''$ per axis~\cite{Gillessen:2004tc}, so these positions are consistent within the systematic uncertainty.

\subsubsection*{Spectral model fitting and systematic uncertainties}
\label{spec_fit}
The spectral model fitting and flux estimation was performed using the forward-folding method which accounts for the energy
dispersion and acceptance of the instrument~\cite{refId0}. The spectrum was fitted with a simple power-law to characterise the measured spectrum and a power-law function with EBL attenuation to characterise the intrinsic spectrum.
Systematic studies were carried out to investigate a presence of curvature in the intrinsic spectra by considering the data from the first night and from night 1 to 3. Within the systematic uncertainties of the analysis and EBL models (see below), no preference for curvature was found over the attenuated power-law model considered.

Instrument-related systematic uncertainties in the spectra were determined by introducing a 10\% shift up and down in the
reconstructed energy of the events and re-fitting. This accounts for inaccuracies in the energy estimation of the reconstructed events seen between the two analysis chains. A 20\% uncertainty in the flux normalisation and 0.09 in photon index were added in quadrature to account for other systematic uncertainties (atmospheric shower interaction models, atmospheric effects, broken pixels in the telescope cameras, etc.)~\cite{HESSCrab}. The choice of model for the EBL absorption was considered a possible source of systematic uncertainty, computed using three different EBL models\cite{Franceschini2008,Dominguez2011,Finke2010} and added in quadrature to the other uncertainties. Overall, the systematic uncertainties are not dominated by the choice of EBL model. Specifically, for the spectrum of the combined three nights of GRB observations by \hess, the instrument-related systematic uncertainties leads to a factor of 2 larger uncertainty on $\photonindex\intr\vhe$ (and $\sim$20\% on $N_0$) compared to the corresponding EBL systematic uncertainty.  For the VHE photon energies, the corresponding EBL wavelength dominating absorption is around a micron, where the corresponding uncertainty of the EBL is more than a factor of 3\cite{HESSEBL}. The spectral fit results for the observations are summarised in Table~~\ref{tab:HESSspectrum}.

\begin{table}
\centering
    \begin{tabular}{ccccc}
    \toprule
    \multicolumn{5}{c}{Power-law model}\\
    \midrule 
    Interval & Time after T$_0$ [s] & $\photonindex\obs\vhe$ & N$_0\times 10^{-12}$ (TeV$^{-1}$cm$^{-2}$s$^{-1}$) & E$_0$ \\ 
    \midrule

    Night 1 & 1.56$\times$10$^4$ -- 2.85$\times$10$^4$ & $2.59\pm 0.09$\,(stat.)\,$\pm 0.23$(syst.) & $13.95\pm 1.05$\,(stat.)\,$\pm 2.92$(syst.) & 0.556 \\
   Night 2 & 9.79$\times$10$^4$ -- 1.15$\times$10$^5$ & $2.46\pm 0.22$\,(stat.)\,$\pm 0.14$(syst.) & $1.22\pm 0.27$\,(stat.)\,$\pm 0.27$(syst.) & 0.741 \\
   Night 1-3 & 1.56$\times$10$^4$ -- 2.01$\times$10$^5$ & $2.59\pm 0.08$\,(stat.)\,$\pm 0.20$(syst.) & $5.01\pm 0.38$\,(stat.)\,$\pm 1.04$(syst.) & 0.583 \\ 
    \midrule 
    \multicolumn{5}{c}{Power-law with EBL attenuation model}\\
    \midrule 
    Interval & Time after T$_0$ [s] & $\photonindex\intr\vhe$ & N$_0\times 10^{-12}$ (TeV$^{-1}$cm$^{-2}$s$^{-1}$) & E$_0$ \\ 
    \midrule

    Night 1 & 1.56$\times$10$^4$ -- 2.85$\times$10$^4$ & $2.06\pm 0.10$\,(stat.)\,$\pm 0.26$(syst.) & $22.67\pm 1.71$\,(stat.)\,$\pm 4.84$(syst.) & 0.556 \\ 
    Night 2 & 9.79$\times$10$^4$ -- 1.15$\times$10$^5$ & $1.86\pm 0.26$\,(stat.)\,$\pm 0.17$(syst.) & $2.31\pm 0.52$\,(stat.)\,$\pm 0.53$(syst.) & 0.741 \\
    Night 1-3 & 1.56$\times$10$^4$ -- 2.01$\times$10$^5$ & $2.07\pm 0.09$\,(stat.)\,$\pm 0.23$(syst.) & $8.34\pm 0.62$\,(stat.)\,$\pm 1.78$(syst.) & 0.583 \\ 
    \bottomrule
    \end{tabular}
    \caption{Results of the spectral model fitting of the \hess observations. Uncertainties in photon index $\photonindex$ and flux normalisation $N_0$ are statistical and systematic in that order (all uncertainties are 1~sigma).}
    \label{tab:HESSspectrum}
\end{table}

\subsubsection*{VHE temporal decay}
\label{VHE_t_decay}

The temporal evolution of the energy flux detected using \hess was fitted with a power-law model of the form 
$F(E,t) = (t/t_{0})^{-\alpha} F(E)$, where $F(E)$ is the energy flux of the intrinsic spectrum integrated in the energy range between $E_{\rm min}$ and $E_{\rm max}$, $F(E_{\rm min},E_{\rm max}) = \int^{E_{\rm max}}_{E_{\rm min}}~E\,dN/dE~dE$. For this analysis, a constant photon index of $\photonindex\intr\vhe$=$2.07\pm0.09$\,(stat.)\,$\pm0.23$\,(syst.) was adopted, in agreement with the spectral fit of the whole \hess observation window (Tab.~\ref{tab:HESSspectrum}).
The observations performed during the first night were split into three clusters to obtain three data points in the \lc, five data points when including the second and third night of observations. Results are summarised in Table~\ref{tab:hess_lightcurve}. Model fitting of the temporal evolution was performed using the least-squares method and results in a temporal decay index $\alpha\vhe = 1.09\pm0.05$ (Pearson's $\chi^2$ = 0.31, dof = 3, for the power-law to fit the data). A search for shorter-time variability was performed in the data of the first night of observations. For this, a binning of 900\,s was used which provides on average a mean significance of 5.6$\sigma$ per data point. No variability was found and the temporal decay index $\alpha=1.33\pm0.41$ is consistent with the value obtained when considering the whole set of observations from night one to three.

\begin{table}[]
    \centering
    \begin{tabular}{ccc}
    \toprule
Interval & Time after T$_0$ [s] & Energy Flux $\times 10^{-11}$ (erg cm$^{-2}$ s$^{-1}$) \\
\midrule
Night 1, cluster 1 & \sct{1.56}{4} -- \sct{1.92}{4} & 4.06 $\pm$ 0.65(stat.) $\pm$ 0.90(syst.)\\
Night 1, cluster 2 & \sct{1.92}{4} -- \sct{2.44}{4} & 3.57 $\pm$ 0.42(stat.) $\pm$ 0.77(syst.)\\
Night 1, cluster 3 & \sct{2.44}{4} -- \sct{2.85}{4} & 2.66 $\pm$ 0.39(stat.) $\pm$ 0.60(syst.)\\
Night 1 & \sct{1.56}{4} -- \sct{2.85}{4} & 3.34 $\pm$ 0.28(stat.) $\pm$ 0.72(syst.)\\
Night 2 & \sct{9.79}{4} -- \sct{1.15}{5} & 0.64 $\pm$ 0.14(stat.) $\pm$ 0.14(syst.)\\
Night 3 & \sct{1.84}{5} -- \sct{2.01}{5} & 0.27 $\pm$ 0.13(stat.) $\pm$ 0.07(syst.)\\

\bottomrule
    \end{tabular}
    \caption{Energy flux evolution of the intrinsic VHE emission during the three consecutive nights used to determine $\alpha\vhe$. 
    The photon index values are assumed to be constant with a value of $2.07~\pm~0.09$ as determined from the joint fitting of nights 1 to 3. The energy flux level of the full Night 1 time interval is not included in the determination of $\alpha\vhe$ and is listed for comparison only.}
    \label{tab:hess_lightcurve}
\end{table}


\subsection*{Extragalactic Background Light absorption}

The effect of the Extragalactic Background Light (EBL) on the VHE emission of the GRBs is visualised in Fig.~\ref{fig:absorption_ebl_int}. Here, in the same energy range, we compare the level of absorption due to the EBL ($e^{-\tau}$) according to the redshift of the diffenent sources: $z=0.0785$ for \grb \cite{2020arXiv200904021H}; $z=0.4245$ for GRB190114C \cite{2019Natur.575..455M}; $z=0.653$ for GRB180720B \cite{2019Natur.575..464A}. The model dependent values of $\tau$ are computed through interpolation of the tabulated values in the original source for the models in \cite{Franceschini2008,Franceschini2017, Dominguez2011} or from tabulation extracted from their implementation in the \textsc{gammapy} package for the models in \cite{Finke2010,Gilmore 2012}.

\subsection*{\emph{Swift} Observations and Data Analysis}

\xrt observations of GRBs are automatically processed by the UK Swift Science Data Centre (UKSSDC) to produce \lcs and time-averaged spectra \cite{2007A&A...469..379E, Evans09}, as well as flux light curves in which spectral evolution is accounted for by means of the hardness ratio\cite{2010A&A...519A.102E}. Because the hardness ratio has larger bins than the light curve, the flux bins in the burst analyser are not truly independent. Therefore, we produced larger time bins using the data taken with the Photon Counting data mode and produced independent spectra for each of these bins using the 'Create time-sliced spectra' online tool,
then fitted these spectra to determine the photon index and flux in each time bin. The spectra were fitted simultaneously so that the column density at the source $N_\mathrm{H,int}$ was tied across all spectra but was free to vary; the spectra were otherwise independent. This approach gives lower time resolution than available from the burst analyser, but gives more precise flux measurements during the  H.E.S.S. observations.

We identified 16 time bins for which these spectra were generated. In the first instance, we generated 11 bins which were chosen to be as short as possible while still containing enough photon counts per interval to constrain the one-sided uncertainty in the photon index to be less than 0.2, or around 10\% (Fig.~\ref{fig:mwl_lightcurve}). The other 5 bins (which overlap the 11 bins just mentioned) were chosen for their coincidence with H.E.S.S. observations. Two intervals were chosen to coincide with the first two nights of \hess observations (there were no XRT observations coincident with the third night), and were used in the multiwavelength spectral modeling (Fig.~\ref{fig:MWL_fit_sync_limit}); the first night was additionally split into three clusters to test for short-term evolution (Fig.~\ref{fig:mwl_lightcurve}).
The 16 spectra were fitted simultaneously in {\sc xspec} 12.10.1 \cite{Arnaud96} with an absorbed power-law model, using standard abundances \cite{Wilms00} and cross-sections \cite{Verner96}. We included two absorbers: a Milky Way foreground (using the {\sc tbabs} model) was fixed at the Galactic value of $N_H=5.60\times10^{20}$~cm$^{-2}$ \cite{Willingale13}, and a host galaxy ({\sc ztbabs}) component fixed to the redshift of the GRB, with $N_\mathrm{H,int}$ free to vary but tied to be the same for all 16 spectra, and found $N_\mathrm{H,int} = (1.16 \pm 0.04) \times 10^{22}$ cm$^{-2}$. 
Uncertainties were determined (at the 1-$\sigma$ level) using the {\sc error} command in {\sc xspec}. Table~\ref{tab:XRTfits} lists the results of the fitting to the XRT time intervals used in the spectral and temporal analyses, and Table~\ref{tab:XRTfits_closedcircles} those used to constrain the X-ray photon index.

\begin{table}[]
    \centering
    \begin{tabular}{cccc}    
    \toprule
    Interval & Time after T$_0$ [s] & $\log_{10}(F_\mathrm{XRT}$) (erg cm$^{-2}$ s$^{-1}$) & $\gamma_\mathrm{XRT}$ \\
    \midrule  
    Night 1\,\,\, & \sct{1.56}{4} -- \sct{2.85}{4} & $-10.03 \pm 0.02$ & $2.03 \pm 0.06$ \\   
    Night 2\footnote{These time intervals were used for calculating the temporal decay index of the X-ray emission.} & \sct{9.79}{4} -- \sct{1.15}{5} & $-10.74 \pm 0.03$ & $2.04 \pm 0.10$ \\   
    \midrule
    Night 1, cluster 1$^a$ & \sct{1.56}{4} -- \sct{1.92}{4} & $-9.85 \pm 0.02$ & $1.93 \pm 0.08$ \\  
    Night 1, cluster 2$^a$ & \sct{1.92}{4} -- \sct{2.44}{4} & $-10.07 \pm 0.03$ & $2.14 \pm 0.10$ \\ 
    Night 1, cluster 3$^a$ & \sct{2.44}{4} -- \sct{2.85}{4} & $-10.14 \pm 0.04$ & $2.12 \pm 0.14$ \\ 
    \bottomrule
    \end{tabular}
    \caption{XRT data coincident with the first two nights of \hess observations (first two rows) are used in the multiwavelength spectral analyses; no XRT data were available coincident with the third night. XRT data coincident with the first night were additionally divided into three shorter time intervals to probe the X-ray evolution (last three rows).}
    \label{tab:XRTfits}
\end{table}

\begin{table}[]
    \centering
    \begin{tabular}{ccc}
    \toprule
    Time after T$_0$ [s] & $\log_{10}(F_\mathrm{XRT}$) (erg cm$^{-2}$ s$^{-1}$) & $\gamma_\mathrm{XRT}$ \\
    \midrule
    \sct{2.86}{2} -- \sct{4.42}{2} & $-9.25 \pm 0.03$ & $1.96 \pm 0.15$ \\
    \sct{4.42}{2} -- \sct{7.25}{2} & $-9.34 \pm 0.03$ & $2.19 \pm 0.12$ \\
    \sct{7.25}{2} -- \sct{1.17}{3} & $-9.16 \pm 0.02$ & $1.92 \pm 0.08$ \\
    \sct{4.90}{3} -- \sct{6.17}{3} & $-9.06 \pm 0.02$ & $1.85 \pm 0.07$ \\
    \sct{6.17}{3} -- \sct{7.49}{3} & $-9.19 \pm 0.02$ & $1.91 \pm 0.07$ \\
    \sct{1.08}{4} -- \sct{1.90}{4} & $-9.67 \pm 0.01$ & $2.00 \pm 0.05$ \\
    \sct{2.28}{4} -- \sct{3.62}{4} & $-10.20 \pm 0.02$ & $2.08 \pm 0.07$ \\
    \sct{3.62}{4} -- \sct{1.79}{5} & $-10.67 \pm 0.02$ & $2.18 \pm 0.05$ \\
    \sct{2.58}{5} -- \sct{3.28}{5} & $-11.10 \pm 0.03$ & $2.27 \pm 0.09$ \\
    \sct{3.28}{5} -- \sct{5.28}{5} & $-11.32 \pm 0.03$ & $2.37 \pm 0.11$ \\
    \sct{5.28}{5} -- \sct{1.56}{6} & $-11.69 \pm 0.03$ & $2.59 \pm 0.10$ \\
    \bottomrule
    \end{tabular}
    \caption{The 11 time intervals used to study the long-term ($t < 10^6$~s) behavior of the X-ray afterglow observed by XRT. The XRT unabsorbed energy flux $F_\mathrm{XRT}$ and the index $\gamma_\mathrm{XRT}$ were determined by fitting a power-law model to the data in each time bin. These data are plotted in Fig.~\ref{fig:mwl_lightcurve}.}
    \label{tab:XRTfits_closedcircles}
\end{table}

\subsubsection*{Temporal behavior}
Using the \xrt energy flux values over the three nights of \hess observations and binned in the way described above,
we derived a time decay index of $\alpha\XRT= 1.07 \pm 0.09$ (Pearson's $\chi^2$ = 2.63, ndof = 2), in agreement with the decay index of the VHE data. Fitting the data in the original time binning provided by \xrt yielded a decay index of $\alpha\XRT =1.03 \pm 0.02$, consistent with the fit to the binned data. 
We investigated also the presence of a break in the power-law model, and found it with a p-value of 0.998 ($\sim$3$\sigma$). However the resulting break time-scale was not well constrained: $(1 \pm 2)\times10^5$ s.
Automated \BAT data are also provided by the UKSSDC, and the BAT data points plotted in Fig.~\ref{fig:mwl_lightcurve} 
were obtained from the Burst Analyser.

We also examined the temporal behaviour of the \xrt data for GRB~180720B and GRB~190114C. To simplify comparisons with GRB~190829A, we considered only the data between $1.56\times10^{4}$ and $2.01\times10^{5}$~s for each GRB and fitted the 0.3--10~keV energy flux data --- obtained from the automated XRT lightcurve repository at the UKSSDC 
--- in their original binning using both power-law and broken power-law models. We find that GRB~180720B is well described by a power-law with an index of $\alpha\XRT= 1.29 \pm 0.02$; for GRB~190114C, we find a preference for a broken power-law, with significance above $5\,\sigma$, finding a break time at $2.7\times10^{4} \pm 1.5\times10^{3}$~s, with pre- and post-break indices of ${\alpha\XRT}_{1} = 2.4 \pm 0.2$ and ${\alpha\XRT}_{2}=1.13 \pm 0.05$. Compared to the temporal decay index for GRB~190829A, these values are slightly more representative of the average decay index during this time, but in general, there is no single characteristic temporal decay index during this time for the three GRBs. We note that we considered only the data within this limited time interval, and the resultant fits would therefore not be expected to reproduce the best-fit lightcurves reported by the automated fitting procedure in the online XRT catalogue.

In order to place the temporal decay of GRB~190829A in context, we also considered the best-fit broken power-law lightcurves for the entire population of GRB afterglows detected by XRT, provided by the online XRT catalogue. For each GRB, we considered the behavior of the best-fit lightcurve within this same time range ($1.56\times10^{4}$ to $2.01\times10^5$~s) and calculated the time-averaged decay index. We found that the average decay index of XRT lightcurves during this time is around 1.4; the minimum value is around 0 and the maximum around 4, and the bulk of the population falls between 0.5 and 2.

\subsubsection*{Fluence and energy calculations}
\begin{figure}[ht!]
    \includegraphics[width=1.0\textwidth]{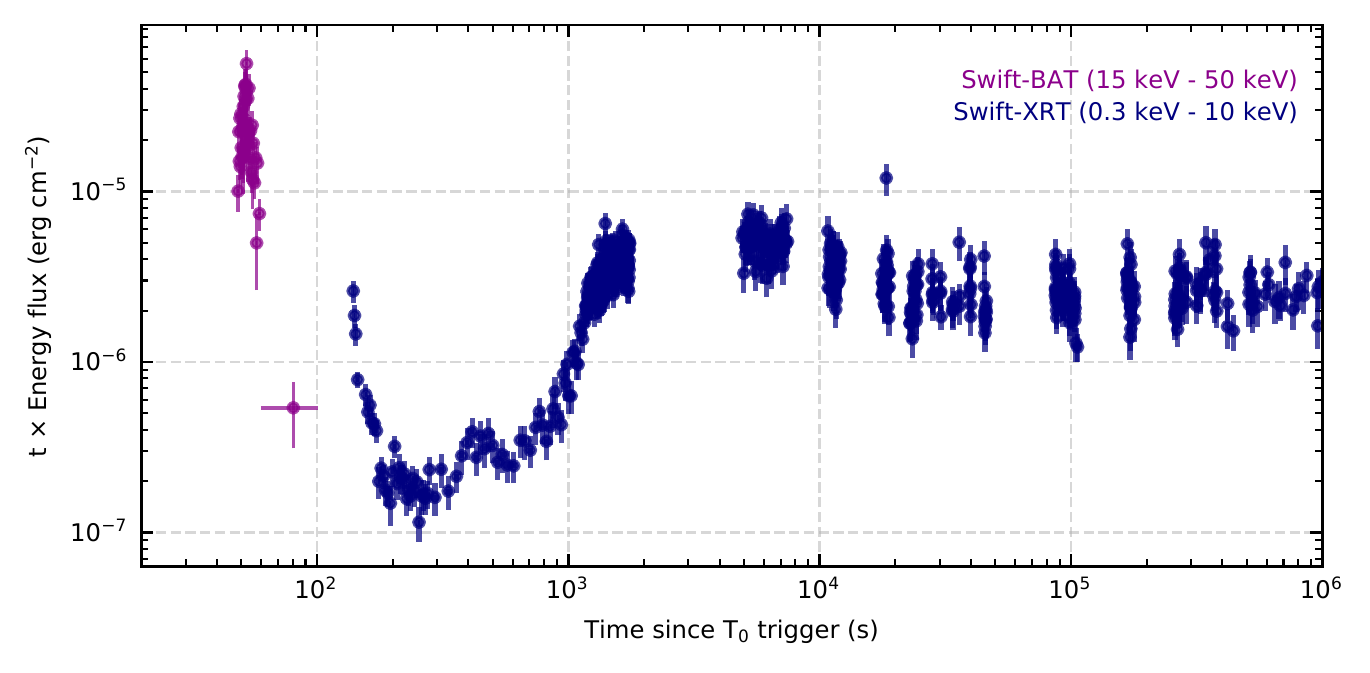}
    \caption{Temporal analys of the X-ray \lc of \grb. The prompt emission detected by the \BAT is shown in purple and the afterglow detected by \xrt is shown in blue. Error bars show 1$\sigma$ uncertainty.}
    \label{fig:fluence_evol}
\end{figure}

We consider the fluence contained within the XRT afterglow \lc, obtained by integrating the energy flux within the instrument energy band over time, from $t_\mathrm{min}$ to $t_\mathrm{max}$, ${\rm fluence} = \int_{\ln t_{\rm min}}^{\ln t_{\rm max}}~t F(t) d\ln t$. By doing this we determine the timescale which dominates the fluence integral. The integrand function is shown in Fig.~\ref{fig:fluence_evol}, namely $tF(t)$. This indicates that for \grb, no specific time interval dominates the fluence. Instead, the fluence is found to increase approximately with the logarithm of the temporal integration range.

To further study the energy release in the XRT band, we calculated the isotropic energy release, $E_\mathrm{iso}$, in the XRT band and compared it to the values of $E_\mathrm{iso}$ as reported by the BAT and GBM. We considered the XRT emission in three time intervals: a) $T_0$ to $T_0 + 87.5$~s, b) $T_0 + 87.5$~s to $T_0 + 1600$~s, and $T_0 + 1600$~s to $T_0 + 10^6$~s. For time interval (a), we extrapolated the time-integrated spectrum reported by the GBM \cite{2019GCN.25575....1L} --- which has better constraints than the BAT given that the GBM's energy range extends to lower energies --- into the XRT energy range. For time interval (b), during which there is good XRT coverage, we calculated the fluence in each time bin by multiplying the bin width by the energy flux, and summed up the binned fluences. For time interval (c), at these times there are gaps in the XRT coverage but the lightcurve decays smoothly, so we fit the energy flux lightcurve with a power-law and integrated the power-law fit to obtain the fluence. (We also tried power-laws with one or two breaks but were unable to obtain a better fit result; note that the automated best-fit lightcurve provided by the UKSSDC 
excludes the data between $T_0 + 939$~s and $1.15\times10^{4}$~s as being a flare and so does not account for all of the fluence.) We found that the fluences in these three time intervals are approximately (a) $5\times10^{-6}$ erg cm$^{-2}$, (b) $2\times10^{-6}$ erg cm$^{-2}$, and (c) $2\times10^{-5}$ erg cm$^{-2}$, for a total energy fluence of around $3\times10^{-5}$ erg cm$^{-2}$. The uncertainty in time interval (c) dominates and is around the 10\% level.

\subsection*{Afterglow phase of GRB}
\label{afterglow}

The dynamics of the forward shock are determined by two factors: the released energy and the density of the circum-burst
medium. In the framework of the one-dimensional dynamical model\cite{1976PhFl...19.1130B}, the conservation of energy yields the following dependence of the forward-shock Lorentz factor on its radius (\(R\)):
\be\label{eq:G}
\Gamma \approx \sqrt{\frac{E\iso}{Mc^2}}\,,
\ee
where \(E\iso\) is equivalent isotropic energy, which can be estimated from the prompt burst radiated energy; c is the speed of light; and \(M\) is the total mass swept-up by the
shock. To obtain the total mass, a specific model for the circum-burst environment must be adopted. Two different
approximations are typically considered: a homogeneous interstellar medium (ISM) of density \(\rho_0\), i.e., \(M\propto\rho_0R^3\) (``ISM'' case); and a constant velocity
stellar wind, i.e., \(M= \dot{m}_* R/v_\infty\) (``wind'' case), where \(\dot{m}_*\) and \(v_\infty\) are  mass-loss rate of the star and terminal velocity of the wind. In the relativistic
regime, the forward shock radius, \(R\simeq ct\), depends only on the time elapsed since the explosion in the progenitor reference frame.

The forward shock propagates towards the observer almost at the light speed, resulting in a strong time compression of the emission detected by a distant observer. The forward shock lags behind the initial burst only by a distance of \(c\dif{t}/(2\Gamma^2)\), where \(\dif{t}\) is a time interval in the progenitor reference frame. Thus, the time delay \(\Delta t=T-T_0\) between the initial burst trigger time, \(T_0\), and the observation time, \(T\), determines the corresponding forward shock radius, \(R\), as 
\be\label{eq:trigger}
\Delta t \approx \int\limits_0^R \frac{\dif{\tilde{R}}}{2c\left(\Gamma(\tilde{R})\right)^2} \,.
\ee
Thus, 
the forward shock radius can be estimated as
\be\label{eq:R}
R\simeq \Gamma^2(R) c \Delta t\left\{
  \begin{matrix}
    8& {\rm for\,\,ISM}\\
    4& {\rm for\,\,wind}\,,
  \end{matrix}
\right.
\ee
where \(R\) is the forward shock radius at the moment of emission.
According to Eqs.~(\ref{eq:G})~and~(\ref{eq:R})  \cite{2019ApJ...880L..27D}, the
forward shock bulk Lorentz factor depends weakly on the burst energy and density
\be
\Gamma\approx\left\{
  \begin{matrix}
    5.6& \left(\kappa\gbm n_0t_{\rm 4h}^3\right)^{\nicefrac{-1}{8}} & {\rm for\,\,ISM}\\
    2.2 &\left(\frac{v_0}{\kappa\gbm\dot{m}_0 t_{\rm 4h}}\right)^{\nicefrac14} &{\rm for\,\,wind}\,.
  \end{matrix}
\right.
\ee
Here and below we normalise the physical parameters as  \( n_0 =\rho_0 /(m_p \rm\,cm^{-3})\) (\(m_p\) is mass of proton), \(\dot{m}_0 = \dot{m}_*/ (10^{-5}\mathrm{M}_\odot\rm\,yr^{-1})\) (\(\mathrm{M}_\odot\) is solar masses), \(v_0 = v_\infty/(10^3\rm\,km\,s^{-1})\), and \(t_{4h} = \Delta t/(4\rm\,hrs)\). The total explosion energy is obtained from the energy radiated in the GBM band with efficiency coefficient \(\kappa\gbm=(2\times10^{50}\rm\,erg)/E\iso\).

At the moment of the production of the gamma-ray emission 
detected with \hess from \grb, the forward shock had reached light-month distances from the progenitor star:
\be
R\approx10^{16}{\rm\,cm}\left\{
  \begin{matrix}
     10\left(\frac{t_{\rm 4h}}{\kappa\gbm n_0}\right)^{\nicefrac14} & {\rm for\,\,ISM}\\
    0.8\left(\frac{v_0 t_{\rm 4h}}{\kappa\gbm \dot{m}_0}\right)^{\nicefrac12} &{\rm for\,\,wind}\,.
  \end{matrix}
\right.
\ee
In the case of the forward shock propagating through the stellar wind, the above solution is valid on scales of stellar wind bubbles. Given that the stellar bubbles can exceed \(10\rm\,pc\), the above solution is likely valid even during the third night (\(t_{\rm 4h}\simeq10\)).  

Here and below we consider two reference frames, the emission production frame associated with the forward shock, and the observer frame. The
quantities in the forward shock frame are indicated with primed symbol, e.g., \(\varepsilon'\), and in the observer frame are plain, e.g.,
\(\varepsilon\). Thermodynamic parameters, like density or internal energy, are always taken in the plasma co-moving
frame, but we nevertheless prime them for clarity. The framework where one considers only
two reference frames, is a simplification. For example, we do not distinguish the observer and the progenitor
frames, because of the small redshift for this GRB, \(z\ll1\). We also assume that the production region moves toward the
observer with the forward shock bulk Lorentz factor. This assumption has two consequences. First, that the forward shock is spherical and therefore different regions of the shock have different velocity directions. Second, that the shocked circum-burst medium moves relativistically away from the observer in the reference frame of the forward shock. For a
weakly magnetised shock, the bulk Lorentz factors of the forward shock and shocked medium differ by a factor of
\(\sqrt{2}\) in the observer frame.

In the reference frame of the forward shock, the flow speeds in the upstream, \(v_{\rm us}\), and downstream, \(v_{\rm ds}\), satisfy the following condition \(v_{\rm ds}v_{\rm us}=v_s^2\) (see, e.g., \cite{Landau6}). Here \(v_s=c/\sqrt{3}\) is sound speed in relativistic gas. In the limit of high Lorentz factor of the upstream flow, \(v_{\rm us}\rightarrow c\), one obtains \(v_{\rm ds}=c/3\). By considering the matter fluxes through the shock, \(v_{\rm ds}\Gamma_{\rm ds} \rho_{\rm ds} = c\Gamma\rho_{\rm us}\), the density of the shocked plasma, \(\rho_{\rm ds}\), can be obtained 
\be
\rho_{\rm ds}' \simeq \sqrt{8}\Gamma \rho_{\rm us}\,,
\ee
where we have accounted for the fact that \(\Gamma_{\rm ds}=1/\sqrt{1-\left(v_{\rm ds}/c\right)^2}=3/\sqrt{8}\).  The upstream density for the forward shock, \(\rho_{\rm us}\), is the density of the circum-burst medium. As the density in shocked plasma should be nearly constant, the thickness of the shocked plasma layer can be estimated as
\be
\Delta R'\approx \frac{M}{4\pi R^2 \rho_{\rm ds}'}\approx\frac{R}{\Gamma}\left\{
  \begin{matrix}
     \frac{1}{3\sqrt{8}} & {\rm for\,\,ISM}\\
    \frac{1}{\sqrt{8}}  &{\rm for\,\,wind}\,.
  \end{matrix}
\right.
\label{emission_shell}
\ee
The energy flux through the shock allows the determination of the internal energy of the shocked plasma. In a relativistic fluid, the energy flux is \((w+p)\Gamma^2v\) (here \(w\) and \(p\) are internal energy and gas pressure, see, e.g., \cite{Landau6}). While in the downstream region the relativistic equation of state, \(p_{\rm ds}=w_{\rm ds}/3\), holds, the upstream is dominated by the rest energy density, \(p_{\rm us}\ll w_{\rm us}\approx\rho_{\rm us}c^2\). Thus, one obtains
\be\label{eq:wds}
w_{\rm ds}'\approx\frac34\rho_{\rm us}c^2\frac{\Gamma^2c}{\Gamma_{\rm ds}^2 v_{\rm ds}}\approx 2\Gamma^2 \rho_{\rm us} c^2\,.
\ee
This defines the strength of the magnetic field in the production region, \(B'\), by its contribution to the downstream internal energy:
\be\label{eq:wb}
w_{\rm B}'=\frac{B'^2}{8\pi}=\eta_{\rm B} w_{\rm ds}'\,,
\ee
where $\eta_{B}$ is the downstream magnetisation.
For the considered models of the circum-burst environment, one obtains:
\be\label{eq:B_co}
B'\approx\left\{
  \begin{matrix}
     1.5{\,\rm G}&\left(\frac{n_0^3\eta_{\rm B}^4}{\kappa\gbm t_{\rm 4h}^3}\right)^{\nicefrac18} & {\rm for\,\,ISM}\\
    43{\,\rm G}&\left(\frac{\dot{m}_0^3\eta_{\rm B}^2\kappa\gbm }{v_0^3 t_{\rm 4h}^3}\right)^{\nicefrac14} &{\rm for\,\,wind}\,.
  \end{matrix}
\right.
\ee
The medium magnetisation is constrained by a strict limit \(\eta_{\rm B}<1\) on the high end, but a broad range of weaker magnetic fields is allowed. 

The characteristic energy flux of the ejecta can be estimated as \(\Gamma^2\rho_{\rm us} c^3\). If a fraction, \(\eta\), of this power is radiated, then the source luminosity, \(L\iso\), can be estimated as
\be\label{eq:l_bol}
L\iso=4\pi R^2 \eta  \Gamma^4\rho_{\rm us} c^3\,.
\ee
The energy density, \(w_{\rm ph}'\), corresponding to this emission in the forward shock frame is
\be\label{eq:wph}
w_{\rm ph}' =\frac{L\iso}{4\pi R^2c \Gamma^2}= \eta \Gamma^2\rho_{\rm us} c^2\,.
\ee
%
The ratio of the synchrotron and Thomson emission components (\(L_{\rm syn}\) and  \(L_{\rm Th}\), respectively) is determined by the ratio of the magnetic to photon target densities, i.e.,
\be\label{eq:Syn_to_IC}
\frac{L_{\rm syn}}{L_{\rm Th}}=\frac{\eta_{\rm B}}{\eta}\,.
\ee
Substituting the bulk Lorentz factor and density to Eq.~\eqref{eq:l_bol} one obtains that
\be\label{eq:lightcurve_on_eta}
L\iso\approx10^{45}\frac{\eta}{\kappa\gbm t_{\rm 4h}}\,{\rm erg\,s^{-1}}\left\{
  \begin{matrix}
     6& {\rm for\,\,ISM}\\
    4&{\rm for\,\,wind}\,.
  \end{matrix}
\right.
\ee
This suggests that any \lc deviation from \(t^{-1}\) dependence implies a change of the radiation efficiency, \(\eta\).  The radiation efficiency can be estimated as the ratio of the radiation cooling time, \(t_{\rm rad}'\) to the forward shock dynamic time:
\be \label{eq:eta_on_time}
\frac\eta{\eta_{e}}\simeq\frac{\int\limits_{E_{e,\rm min}'}^{E_{e,\rm max}'}q(\tilde{E})\tilde{E}\min(t_{\rm d}'/t_{\rm rad}',1)\dif{\tilde{E}}} {\int\limits_{E_{e,\rm min}'}^{E_{e,\rm max}'}q(\tilde{E})\tilde{E}\dif{\tilde{E}}}\,,
\ee
where \(q\) is the injection spectrum; \(E_{e,\rm max/min}'\) is the energy range in which non-thermal particles are injected in the co-moving frame;  \(\eta_{e}\) is fraction of energy transferred to non-thermal electrons;   \(t_d'=R/\Gamma c\) is the dynamic time. If inverse Compton (IC) scattering proceeds in the Thomson regime, then the radiative cooling time can be expressed as \(t_{\rm rad}'=t_{\rm syn}'/\kappa\), where \(t_{\rm syn}'\simeq 400 B_G'^{-2}E_{\rm TeV}'^{-1}\rm\,s\) is synchrotron cooling time and the dimensionless parameter \(\kappa=1 + w_{\rm ph}'/w_{\rm B}'\) accounts for {\rm additional} IC cooling.   Equation \eqref{eq:eta_on_time} depends on the injection spectrum and does not allow any simple analytical reduction. Adopting a simple power-law injection spectrum, \(q \propto E_{e}'^{-\beta_{\rm inj}}\) (here \(\beta_{\rm inj}\) is the power-law index), the minimum energy of injected non-thermal particles  \cite{Sari1998} can be estimated as
\be \label{eq:minimum_energy}
E_{e,\rm min}'\simeq\frac{(\beta_{\rm inj}-2)}{(\beta_{\rm inj}-1)}\Gamma m_p c^2\eta_e=\left\{
  \begin{matrix}
    0.9\eta_e{\rm\,GeV} \left({\kappa\gbm t_{4h}^3n_0}\right)^{\nicefrac{-1}{8}}      &{\rm for\,ISM}\,,\\
    0.3\eta_e{\rm\,GeV} \left(\frac{v_0}{\kappa\gbm t_{4h}\dot{m}_0}\right)^{\nicefrac{1}{4}} &{\rm for\,wind}\,.\\
  \end{matrix}
  \right.
\ee
(Here and below for the sake of brevity we show analytic expression only for the case when \(\beta_{\rm inj}\neq1\), \(2\), or \(3\); for numerical estimates we adopt \(\beta_{\rm inj}=2.2\))
Then one obtains that 
\be \label{eq:eta_on_time_2}
\frac{\eta}{\eta_e}=\left\{
  \begin{matrix}
    \frac1{3-\beta_{\rm inj}}\left(\frac{E_{e,\rm min}'}{E_{e,\rm br}'}\right)^{\beta_{\rm inj}-2}-\frac{\beta_{\rm inj}-2}{3-\beta_{\rm inj}}\frac{E_{e,\rm min}'}{E_{e,\rm br}'}&{\rm if\quad} E_{e,\rm br}'>E_{e,\rm min}'\,,\\
    1&{\rm if\quad} E_{e,\rm br}'\leq E_{e,\rm min}'\,,
  \end{matrix}
  \right.
  \ee
where \(E_{e,\rm br}'\) is cooling break energy, which is determined by the adopted model for the circumburst medium
\be\label{eq:E_br_co}
E_{e,\rm br}'\approx
\left\{
    \begin{matrix}
      \frac{0.2{\rm\,GeV}}{\eta_B\kappa} \left( \frac{\kappa\gbm^3t_{4h}}{3n_0^5}\right)^{1/8}     & \rm for\,ISM\,,\\
      \frac{1{\rm\,MeV}}{\eta_B\kappa} \left(\frac{t_{4h}^3v_0^5}{\kappa\gbm \dot{m}_0^5}\right)^{1/4}& \rm for\,wind\,.\\
    \end{matrix}
\right.
\ee
For the expected properties of \grb (in particular \(\beta_{\rm inj}\sim2\)) and epoch of H.E.S.S., \(t>4\rm\,hrs\), it is expected that the radiation efficiency should saturate close to \(\eta_e\).
We therefore conclude that the X- and gamma-ray \lcs suggest that \(\eta_e\), \(\eta_B\), and \(\eta\) all remained approximately constant during the HESS observation time interval \(4{\rm\,hrs}<\Delta t<56{\rm\,hrs}\).

\subsection*{Multiwavelength Modelling}
\label{modelling}

Adopting the afterglow theory outline above, we developed a numerical modelling tool to interpret the observational data. 
We utilised the python software \textsc{naima} \cite{naima} to model the multiwavelength (MWL) observational data, as well as the full differential cross-sections for the interaction processes. 

Assuming the relations in equations \ref{eq:G}-\ref{eq:R} to define the physical environment of the GRB, we adopted the solutions for the ISM scenario and assumed $E\iso=2\times10^{50}$ erg, $n_0 = 1\,{\rm cm}^{-3}$. The modelling made use of the data coming only from \Swift-XRT, \lat and \hess. Studies at optical and radio wavelengths \cite{2020MNRAS.496.3326R, 2020arXiv200311252F} motivate the choice of using the ISM scenario for the modelling; investigations of the optical emission have been performed for other GRBs \cite{2020arXiv200904021H,2020ApJ...898...42C}, but we preferred to focus only on the X- and gamma-ray datasets.
The model does not have any time dependent implementation and we assumed that the parent electron distribution is a broken power-law with a high energy exponential cutoff described by:
\begin{equation}
f(E') = \exp\left(-\frac{E'}{E_{e,\rm cut}'} \right)\left \{
                     \begin{array}{ll}
                       A (E' / E_s') ^ {-\beta_1}    & : E' < E_{e,\rm br}' \\
                       A (E_{e,\rm br}'/E_s') ^ {\beta_2-\beta_1}
                            (E' / E_s') ^ {-\beta_2} & : E' > E_{e,\rm br}'\\
                     \end{array}
                   \right.
\label{eq:bplec}
\end{equation}
where $A$ is the normalisation parameter in units of 1/eV, $E_s'$ is the scale energy, $E_{e,\rm cut}'$ is the cutoff energy, $E_{e,\rm br}'$ the break energy and $\beta_1$ and $\beta_2$ the indices for the power-law below and above the break energy. The minimum energy of the electron distribution and the normalisation parameter were derived as a combination of the other parameters of the function and $\eta_e$ that in the model is defined as the ratio between $w_{\rm {ds}}'$ (equation~\ref{eq:wds}) and the total energy density of the non-thermal electrons close to the forward shock. The expression for the minimum injection energy is a generalisation of Equation~\ref{eq:minimum_energy}. Once the minimum injection and the general shape of the electron distribution were defined, the normalisation $A$ was derived from the ratio between the total energy contained in the electron spectrum with unitary normalisation and the total energy injected given by $\eta_e$.

The model fitting was carried out using two approaches. In the first approach, the low energy component (up to $\sim$MeV energies) is described by synchrotron emission, with the electron spectral index inferred from the photon index of the \xrt data. In a data-driven methodology, a general constraint was obtained on the magnetic field $B'$, utilising the estimated cooling time in the downstream region. Assuming that the break in the electron distribution is a cooling break, the corresponding break in the synchrotron spectrum located below the energy range tracked by \xrt, the cooling time of the electrons at the energy of the spectral break corresponds roughly to the age of the system. We therefore determined that the magnetic field was within an order of magnitude of 1~Gauss during the first night of observation by \hess

To construct an SSC model in \textsc{naima}, a computation of the number density was required. The assumed geometry is that of an expanding thin shell. The number density of photons in the emission region has been approximated as $(\dif[E',t']{n'})/(4\pi c R^2)$, where $\dif[E',t']{n'}$ is the synchrotron spectrum in the co-moving reference frame. This same distribution of synchrotron photons was then used also to compute the contribution of the photon-photon absorption.

The multiwavelength data for night 1 and night 2 were fitted separately using the Markov-chain Monte Carlo (MCMC) approach implemented in \textsc{naima}, leaving as free parameters $\eta_e$, $E_{e,\rm br}'$, $\beta_2$, $E_{e,\rm cut}'$ and the intensity of the magnetic field $B'$. All parameters were varied on a logarithmic scale (except $\beta_2$) for a more efficient exploration of the parameter space.
We used 128 parallel walkers with a burn in phase of 100 steps and a following phase to compute the distribution of another 200 steps. The priors on the parameters for the Bayesian fitting were chosen with a uniform distribution with sufficiently broad boundaries so as not to affect the fitting. The chain starting parameter vector was derived from a maximum likelihood fit. The only exception is for the parameter $\beta_2$ for which the prior has a Gaussian distribution centred on the value that can be computed from the \xrt datapoints and a width extracted from the uncertainty of the \xrt photon index.

Through the consideration of physical arguments, we implement additional constraints on the model parameters: because we assume synchrotron cooling, we expect that $\beta_2-\beta_1 = 1$; the position of the cooling break is roughly in the position where the cooling time for electrons of that energy is equivalent to the age of the system (after correcting for Lorentz time-delay); the cut-off energy has an upper boundary so as not to violate the maximum synchrotron energy limit \cite{1983MNRAS.205..593G}.
The implementation of these constraints was done in the construction of the probability prior function. For the break energy parameter, this was allowed to vary freely between the minimum injection energy of the electrons and the energy required to have the synchrotron characteristic energy equal to a third of the lowest energy \xrt spectral point. This setup ensured that the break energy in the photon spectrum occurs just below the energy of the \xrt points. 

For the magnetic field, a uniform prior between $10^{-3}$ and $10^1$~G (in logarithmic space) was adopted. The constraint linking the break energy and the age of the system was obtained through a Gaussian prior on the cooling time at the break with the Lorentz transformed age of the system as a mean and half of this value as width.
Because the \hess and \xrt spectral analyses were performed via forward-folding techniques, the fitting was performed using data directly derived from the functional power-law spectra adopted. The results of the fits with this configuration are shown in Fig.~\ref{fig:MWL_fit_sync_limit} and final values of the parameters are reported in Table~\ref{tab:par_fullprior}. This model requires a high value for the energy in non-thermal electrons, with large values for $\eta_{e}$. The range of magnetic field values found, correspond to $\eta_B= 0.077$ in the first night and $\eta_B = 0.059$ in the second night.

\begin{table}[ht]
    \centering
    \begin{tabular}{l|c|c}
    \toprule
        Parameter         & Night 1 & Night 2\\[2pt]
    \midrule
    $\log_{10}(\eta_e)$                 & $-0.04\pm0.02$              & $-0.04^{+0.03}_{-0.04}$\\[2pt]
    $\log_{10}(E_{e,\rm br}'/\rm TeV)$  & $-1.513^{+0.007}_{-0.015}$  & $-1.31^{+0.09}_{-0.16}$\\[2pt]
    $\beta_2$                           &  $3.15 \pm 0.03$            & $3.06^{+0.03}_{-0.04}$\\[2pt]
    $\log_{10}(E_{e,\rm cut}'/\rm TeV)$ & $1.7\pm0.2$                 & $1.6^{+0.4}_{-0.5}$\\[2pt]
    $\log_{10}(B'/\rm G)$               & $-0.448^{+0.008}_{-0.004}$  & $-0.77^{+0.08}_{-0.04}$\\[2pt]
    \midrule
    \midrule
    \makecell{Photon index \\ in \hess energy range} & $2.781\pm0.006$ & $2.6 \pm 0.1$\\[2pt]
    \makecell{Energy flux \\ in \hess energy range} & $1.88\times10^{-11}$ erg/cm$^2$/s & $3.07\times10^{-11}$ erg/cm$^2$/s\\
    \bottomrule
    \end{tabular}
    \caption{SSC results with a synchrotron maximum energy constraint imposed. Final parameters values for night 1 and night 2 as in Fig.~\ref{fig:MWL_fit_sync_limit}. The parameter names are as in equation~\ref{eq:bplec}. The best fitting values were computed as the median of the posterior probability distribution, while the associated uncertainties are the 16$^{th}$ and 84$^{th}$ percentiles. The last row reports the photon index of a power-law function fitted to the model in the \hess energy range and the integrated energy flux.}
    \label{tab:par_fullprior}
\end{table}

In this scenario, removing the constraint on the break energy dictated by the age of the system allows the production of a harder electron spectrum. However, the effect improves only very weakly the agreement with the \hess data.

The model fitting performed without imposing a maximum energy of the synchrotron radiation spectrum instead assumed only uniform priors (increasing the upper bound of the magnetic field value up to $10^{3}$~G), without accounting for extra limitations on the values of the parameters, with the exception of the break energy that which was still constrained to produce a break below the \xrt energy range. This setup was chosen so as to allow further exploration of the parameter space. In this case, except for their lower limits, the cutoff energy and the magnetic field intensity were not constrained, with the required magnetic field exceeding the Gauss level. The results from this exploration are shown in Fig.~\ref{fig:MWL_fit_sync_limit} and the parameters are reported in Table~\ref{tab:par_nocutoff}. Due to the loose constraints on some of the parameters, a $2\sigma$ lower limit was determined. For this model, the value of the maximum loglikelihood is of the order $>10^5$ times smaller than that obtained when imposing the maximum energy constraint on the synchrotron spectrum, pointing towards a statistical preference of this model well above 5 standard deviations.
The fit obtained with the synchrotron dominant component found the Inverse Compton component to be highly subdominant. This is reflected in the best-fitting values of the model parameters. For both nights, the value of $\eta_e$ is lower than $10^{-1}$, and the corresponding magnetic field strength is at the level of 10~G, implying $\eta_B \approx 1$. 

\begin{table}[]
    \centering
    \begin{tabular}{l|c|c}
    \toprule
        Parameter         & Night 1 & Night 2\\[2pt]
    \midrule
    $\log_{10}(\eta_e)$                 & $-1.6^{+0.6}_{-1.3}$  & $-1.1^{+0.6}_{-1.3}$\\[2pt]
    $\log_{10}(E_{e,\rm br}'/\rm TeV)$  & $-2.3^{+0.3}_{-0.7}$  & $-2.3^{+0.6}_{-1.2}$\\[2pt]
    $\beta_2$                           & $3.074 \pm 0.006$     & $3.083 \pm 0.018$\\[2pt]
    $\log_{10}(E_{e,\rm cut}'/\rm TeV)$ & $>3.8$                & $>3.5$\\[2pt]
    $\log_{10}(B'/\rm G)$               & $>0.09$               & $>-0.4$\\[2pt]
    \midrule
    \midrule
    \makecell{Photon index \\ in \hess energy range} & $2.151\pm0.008$ & $2.2 \pm 0.1$\\[2pt]
    \makecell{Energy flux \\ in \hess energy range} & $3.07\times10^{-11}$ erg/cm$^2$/s & $4.68\times10^{-12}$ erg/cm$^2$/s\\
    \bottomrule
    \end{tabular}
    \caption{SSC results without a synchrotron maximum energy constraint imposed. Final parameters values for night 1 and night 2 as for the orange envelope in Fig.~\ref{fig:MWL_fit_sync_limit}. The parameter names are as in equation~\ref{eq:bplec}. Cut-off energies and magnetic field intensities are given as $2\sigma$ limits based on the posterior probability distributions. The last row reports the photon index of a power-law function fitted to the model in the \hess energy range and the integrated energy flux.}
    \label{tab:par_nocutoff}
\end{table}

The effect of photon-photon absorption on the high-energy GRB emission was also taken into account assuming a size of the region given by $R/(9\Gamma)$. With the same region both emitting and absorbing radiation, the effect of absorption on the observed flux is thus 
\begin{equation}
F^{\rm obs} = F^{\rm int}\frac{1}{\tau} (1-e^{-\tau}).
\label{eq:gammagammaabs}
\end{equation}
We assume here that emitting electrons are homogeneously distributed in the shell. This approximation may overestimate the attenuation if the emitting particles are localised close to the FS, or underestimate it if they are accumulated deeper in the shell.

\begin{figure}
    \centering
    \includegraphics[scale=0.75]{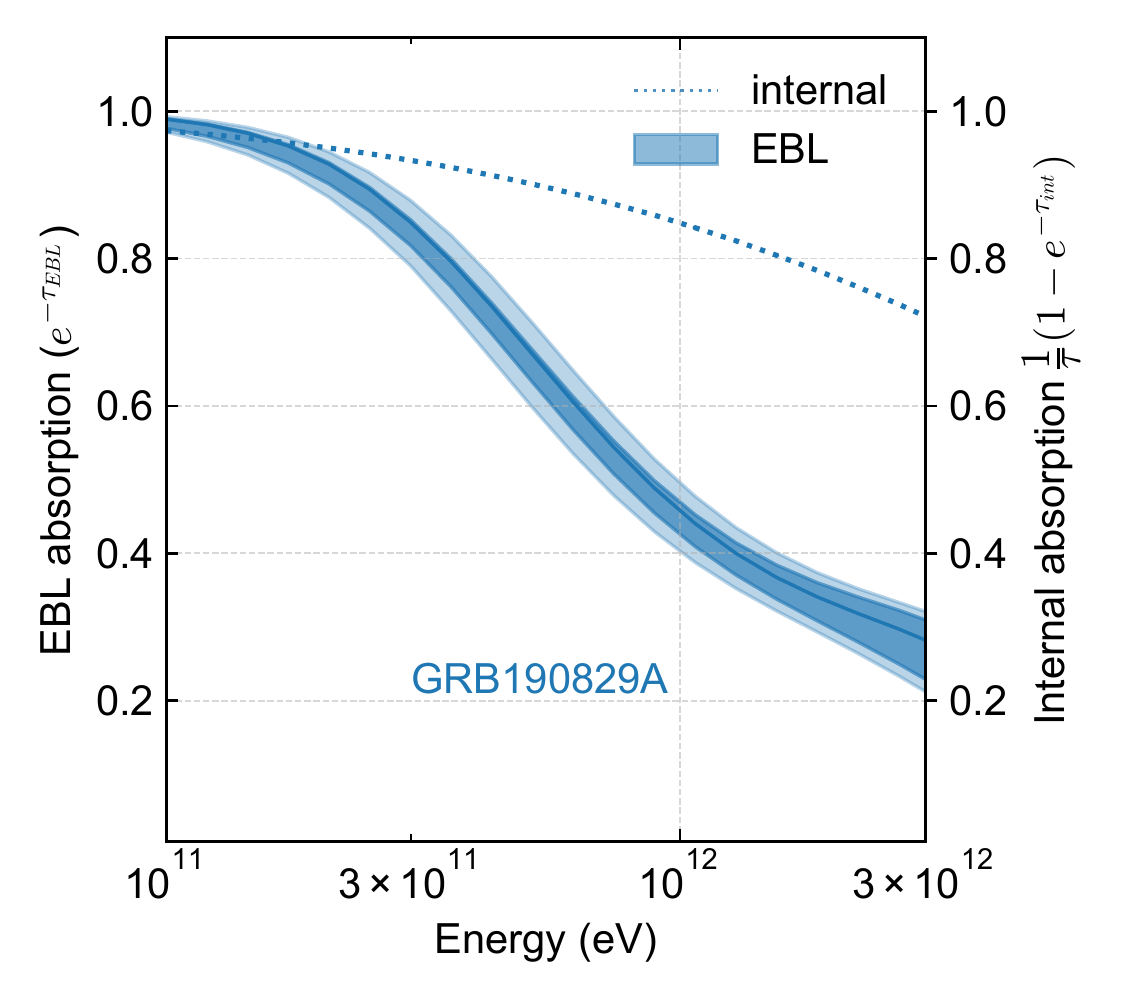}
    \caption{Comparison between internal and EBL absorption. The shaded areas and solid line represent the amount of EBL absorption for \grb using the same notation used in Fig.~\ref{fig:absorption_ebl_int}. The dotted line represents the contribute of the internal absorption derived from the best fit SSC model of the first night. The labels on the y-axes highlight the difference in the calculation of the absorption factor.}
    \label{fig:int_ebl_grb}
\end{figure}

The level of internal absorption can be compared to that due to the EBL. This is presented in Fig.~\ref{fig:int_ebl_grb} where the effect of the EBL is shown together with that of the internal absorption derived from the best-fit SSC model for the first night of observation. While for energies up to 200 GeV the two values are comparable, the contribution of the internal absorption remains subdominant for higher energies, becoming more than a factor of 2 smaller than the EBL absorption at 1 TeV.

The model spectral emission result was calculated in the frame of the GRB. The curves of the resultant spectral energy distribution produced were transformed into the observer frame by multiplying the emission by a factor $\Gamma^2$ (under the assumption that the Doppler factor $D\sim\Gamma$) after transforming the energy axis by a factor $\Gamma/(1+z)$.

In simple SSC models the onset of the Klein-Nishina effect in spectral energy distributions is often attributed to the energy of electrons which up-scatter in the Klein-Nishina regime the emission produced by electrons at the cooling break \cite{2001ApJ...548..787S,2020arXiv200311252F}. Combining equations (\ref{eq:B_co}) and (\ref{eq:E_br_co}) shows that this corresponds to an electron energy of
\begin{equation}
    E_{e,\textsc{kn}}'\approx\frac{m_e^2c^4}{\nicefrac{(hc)}{(2m_e^3c^6)}E_{e, \rm br}'^2 eB'}\approx\left\{
    \begin{matrix}
         14\, \eta_B^{3/2}\kappa^2\kappa\gbm^{-5/8}n_0^{7/8}t_{4h}^{1/8}{\rm\,TeV}     & \rm for\,ISM\,,\\
      1\, \eta_B^{3/2}\kappa^2 \left(\frac{\kappa\gbm\dot{m}_0^7}{ t_{4h}^3v_0^7}\right)^{1/4}{\rm\,TeV}& \rm for\,wind\,.\\
    \end{matrix}
    \right.
\end{equation} 
Under this simplified assumption, the Klein-Nishina effect may seem to be unimportant for the \hess energy band when even a moderate Doppler boosting is accounted for. However,  this approach is justified only in the case of a steep spectrum target photon spectrum above the break. In the case of a hard target field, (eg. with photon index $\gamma \sim 2$), as for \grb, the Klein-Nishina effect appears at lower energies, as shown in Fig.~\ref{fig:MWL_fit_sync_limit} (see also \cite{2009ApJ...703..675N,2019ApJ...880L..27D}). The same problem concerns the simplified discussion of internal absorption \cite{2020arXiv200311252F}. Given the measurement of \grb spectra with \hess, modelling of this emission requires the treatment of both the Klein-Nishina effect and internal absorption, in particular the convolution of the IC and pair production cross sections over the spectrum of target photons.

\begin{figure}[ht]
    \centering
    \includegraphics[width=4.75in]{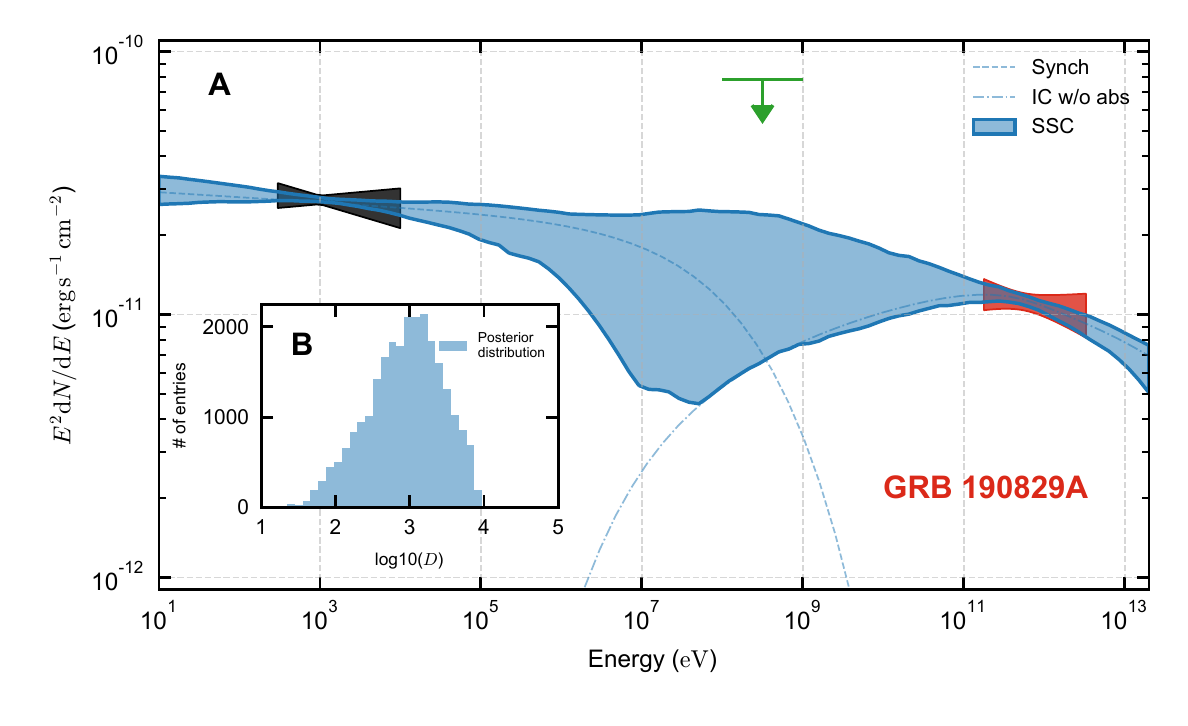}
    \caption{Synchrotron-Self Compton model with additional Doppler boosting. Panel A shows the SSC model fitted to the data of the first night of the \hess observation with the Doppler boosting factor $D$ left as a free parameter. The model lines are obtained for a value $D = 955$. Panel B shows the posterior probability distribution of $\log_{10}(D)$. The colour scheme is the same as in Fig.~\ref{fig:MWL_fit_sync_limit}.}
    \label{fig:freedopp}
\end{figure}

An additional model fitting was also performed in which the Doppler boosting factor is left free to vary instead of being fixed to $\Gamma$. In this case, the prior for the Doppler factor (treated as a logarithmic quantity, as the other parameters) was chosen to be a wide uniform one from a minimum value of $\Gamma$, up to  $10^5$. To obtain the final fit in this case, the prior on the age of the system was removed. The results are shown as a spectral energy distribution (SED) (Fig.~\ref{fig:freedopp}A) and posterior probability distribution of the $\log_{10}$ of the Doppler factor (Fig.~\ref{fig:freedopp}B).

\begin{figure}
    \centering
    \includegraphics{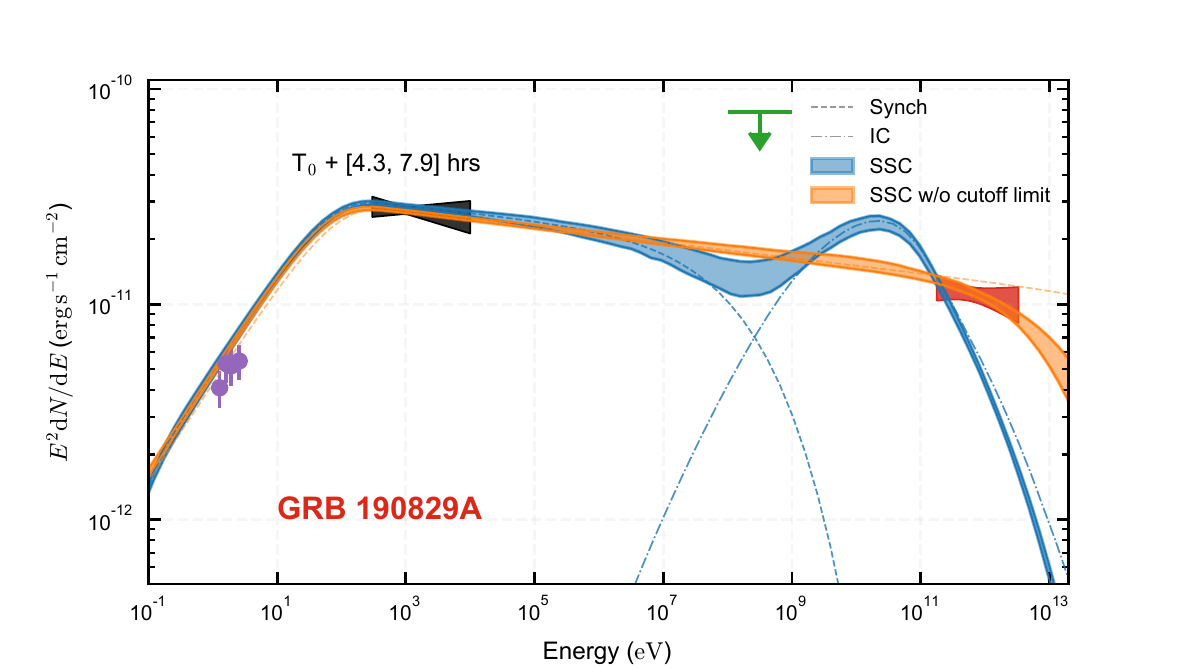}
    \caption{Synchrotron-Self Compton and Synchrotron dominated fits with the optical data also included. The colour scheme is the same as in Fig.~\ref{fig:MWL_fit_sync_limit}. The purple points indicate the optical data \cite{2020arXiv200904021H}.}
    \label{fig:added_optical}
\end{figure}

As a robustness test, we extracted the optical data \cite{2020arXiv200904021H} for the time interval corresponding to the \hess observation of the first night using the interpolation given by those authors. Given that these are synthetic points, extrapolated between observations and ignore the uncertainty on the dust distribution function, we have also assumed an uncertainty equivalent to 20\% of the quoted flux. These additional constraining data were also factored into the MCMC routine. The result of the additional fits are shown in Fig.~\ref{fig:added_optical} and Tab.~\ref{tab:par_add_optical}. By comparing these results with those shown in Fig.~\ref{fig:MWL_fit_sync_limit}, it is apparent that the general shape of the model remains unchanged by the additional consideration of this dataset, and consequently that the discrepancy between the TeV data and the one-zone SSC model remains. The result for the synchrotron-dominated fit model also remain unchanged. A similar analysis for the second night, cannot be performed, due to the presence of features in the optical spectrum which appear related to a thermal component \cite{2020arXiv200904021H}, not taken into account by our pure non-thermal emission model. The radio observations are more difficult to treat given the need to disentangle the forward and the reverse shock (FS and RS respectively). However, the estimate of the FS contribution \cite{2020MNRAS.496.3326R}, also appears to match the level predicted by our best-fitting model.

\begin{table}[]
    \centering
    \begin{tabular}{l|c|c}
    \toprule
        Parameter         & \makecell{Night 1 \\ SSC} & \makecell{Night 2 \\ SSC no limit}\\[2pt]
    \midrule
    $\log_{10}(\eta_e)$                 & $-1.0^{+0.01}_{-0.02}$  & $-2.0^{+0.9}_{-1.5}$\\[2pt]
    $\log_{10}(E_{e,\rm br}'/\rm TeV)$  & $-1.508^{+0.004}_{-0.009}$  & $-2.2^{+0.3}_{-0.6}$\\[2pt]
    $\beta_2$                           & $3.065^{+0.014}_{-0.020}$   & $3.073^{+0.005}_{-0.007}$\\[2pt]
    $\log_{10}(E_{e,\rm cut}'/\rm TeV)$ & $1.75^{0.14}_{0.20}$    & $>3.34$\\[2pt]
    $\log_{10}(B'/\rm G)$               & $-0.450^{0.004}_{0.002}$               & $>0.10$\\[2pt]
    \bottomrule
    \end{tabular}
    \caption{SSC results for night 1 including optical data. Final parameters values for night 1 for the models in Fig.~\ref{fig:added_optical}. The parameter names are as in equation~\ref{eq:bplec}. Cut-off energy and magnetic field intensity for the best-fit model without maximum energy for the synchrotron emission are given as $2\sigma$ limits based on the posterior probability distributions}
    \label{tab:par_add_optical}
\end{table}

\subsection*{Spectrum with hardening}\label{nonpl_spectrum}
For a bulk Lorentz factor expected for the later afterglow phase, the spectral modeling of GRB190829A with one-zone SSC models fails to reproduce the hard TeV spectrum. These difficulties may be resolved if the part of the electron spectrum (around \(E_e'\sim E'_e{}\vhe\)), which is responsible for the VHE has a power-law slope \(\beta\sim2.5\). In the co-moving frame the hardening occurs approximately between \(10\)~GeV and \(1\)~TeV. On the other hand the emission registered in the X-ray band has a usual photon index of \(\gamma\sim2\), thus it is produced by electrons with \(E_e'\sim E'_e{}\XRT\) following a power-law spectrum with index of \(\beta\sim3\). There are two possibilities, the X-ray emission can be produced either by electrons below the hard spectrum range, e.g., by electrons with \(E'\sim3\rm\,GeV\) or above, \(E'\sim3\rm\,TeV\). The X-ray emission appears in the co-moving frame at \(\sim100\rm\,eV\), this scenario would require magnetic fields of either \(B'\sim100\rm\, G\) or \(B'\sim10^{-4}\rm\, G\), respectively.  Using Eq.~\eqref{eq:B_co}, we obtain that for ISM circumburst medium
\begin{equation}
\frac{n_0^3\eta_B^4}{\kappa\gbm}\sim\left\{
  \begin{matrix}
    10^{16}&{\rm for}& E_{e}{}\XRT<E_{e}{}\vhe\,,\\
    10^{-32}&{\rm for}& E_{e}{}\XRT>E_{e}{}\vhe\,.\\
  \end{matrix}
\right.
\end{equation}
On the other hand from Eqs.~\eqref{eq:Syn_to_IC} and \eqref{eq:lightcurve_on_eta} we see that \(\eta\sim\eta_B\) and \(\eta\sim \kappa\gbm\). Thus, the above equation approximately reduces to
\begin{equation}
\frac{n_0\eta_B}{\kappa\gbm}\sim\left\{
  \begin{matrix}
    10^{5}&{\rm for}& E_{e}{}\XRT<E_{e}{}\vhe\,,\\
    10^{-11}&{\rm for}& E_{e}{}\XRT>E_{e}{}\vhe\,.\\
  \end{matrix}
\right.
\end{equation}
Given the requirement that \(\eta_B<1\), the strong magnetic field case would imply a very dense circumburst medium of \(n_0\sim10^5\). In contrast, the weak magnetic field case implies a very diluted medium, \(n_0\sim10^{-5}\), if we adopt \(\eta_B\sim10^{-5}\) (which in turn implies that \(E\iso\sim10^{55}\rm\,erg\)). Both of these extreme regimes go beyond the typical parameter space. 

\subsubsection*{Klein-Nishina losses feedback}\label{KN_feedback}
In the fast cooling regime the feedback of the dominant loss mechanism on the particle distribution needs to be taken into account. For high-energy particles,  only the synchrotron or IC cooling might be important. While the Thomson  and synchrotron losses' depend on electron energy as \(\dot{E}\propto E^2\), the Klein-Nishina losses have a different dependence on electron energy, \(\dot{E}\propto\ln E\) (for \(E\gg m_e^2c^4/\varepsilon\), where \(\varepsilon\) is target photon energy, see, e.g., \cite{1985Ap.....23..650A,2014ApJ...783..100K}). Therefore, for the same injected particle distribution, the electron spectrum formed in environments in which cooling is dominated by photon scattering, should depend on the scattering regime. For example, if we assume that the injected distribution obeys a pure power-law dependence, \(\propto E^{-\beta_{\rm inj}}\), the particle spectrum formed under the dominant Thomson losses is \(\propto E^{-(\beta_{\rm inj}+1)}\). The synchrotron and Thomson spectra have the photon index of \({(\beta_{\rm inj}+1)/2}\). For a typical injection spectrum of \(\beta_{\rm inj}\approx2\), these spectra are ``flat'' (photon index of \(\approx2\)). Under the dominant Klein-Nishina losses, the situation is very different. Ignoring the logarithmic factors, Klein-Nishina losses hardens the injected distribution, \(\propto E^{-(\beta_{\rm inj}-1)}\). Such a hard electron distribution produces a gamma-ray spectrum with the photon index of \(\approx \beta_{\rm inj}\). The Klein-Nishina cooling effect has a  much stronger impact on the synchrotron spectrum (e.g., \cite{2005AIPC..745..359K}). The synchrotron spectrum produced by electrons cooled via Klein-Nishina losses is very hard, with photon index of \(\beta_{\rm inj}/2\). For example, if the injection spectrum is close to the conventional value, \(\beta_{\rm inj}\approx2\), then the gamma-ray slope is \(\approx2\). The synchrotron spectrum, in this case, appears to have a ``step-like'' shape: at low energies, where the IC scattering proceeds in the Thomson regime, the slope is usual, \(\approx2\), at higher energies, where the Klein-Nishina effect is important, the synchrotron spectrum hardens approaching the limiting photon index of \(\approx1\). At highest energies, where the Klein-Nishina losses drop below the synchrotron losses, the spectrum recovers to the usual slope of \(\approx2\).

To illustrate these effects we consider a simple radiative model that accounts for the Klein-Nishina effect both for the energy losses and emission spectra. We consider the energy range, where the electron distribution is formed in the fast cooling regime, i.e., one can use stationary equation to describe the electron distribution
\be
\frac{\partial \dot{E} n}{\partial E} = q(E)\,.
\ee
Here \(q\propto E^{-\beta{\rm inj}}\) 
is the injection distribution. For simulations we adopted \(\beta_{\rm inj}=2\) and the injected power of \(6\times10^{44}\rm\,erg\,s^{-1}\). The energy losses account for the synchrotron and IC channels:
\be
\dot{E}=\dot{E}_{\rm syn}+\dot{E}_{  \rm ic}\,,
\ee
where IC scattering proceeds on the synchrotron photons:
\be
\dot{E}_{  \rm ic} = \int \dif{\varepsilon} \dif[\varepsilon,V]{N_{\rm syn}} \dot{E}_{0,\rm ic}(E,\varepsilon)\,.
\ee
Here \(\dif[\varepsilon,V]{N_{\rm syn}}\) is the number density of the synchrotron photons, and \(\dot{E}_{0,\rm ic}\) is energy losses of a electron with energy \(E\) up-scattering a target photon with energy \(\varepsilon\). Despite its simplicity this radiation model is non-linear and does not allow an analytical treatment.

The density of the synchrotron photons depends on the size of the production region \(R\), magnetic field strength \(B\), and the total energy injected to the system per unit of time is \(L_0\). The Compton dominance can be estimated as (see, e.g., \cite{2001ApJ...548..787S})
\be
Y = \frac{-1+\sqrt{1+\frac{18 L_0}{B^2 R^2 c}}}2\,.
\ee

\begin{figure}[ht]
  \centering
  \includegraphics[width=2.75in]{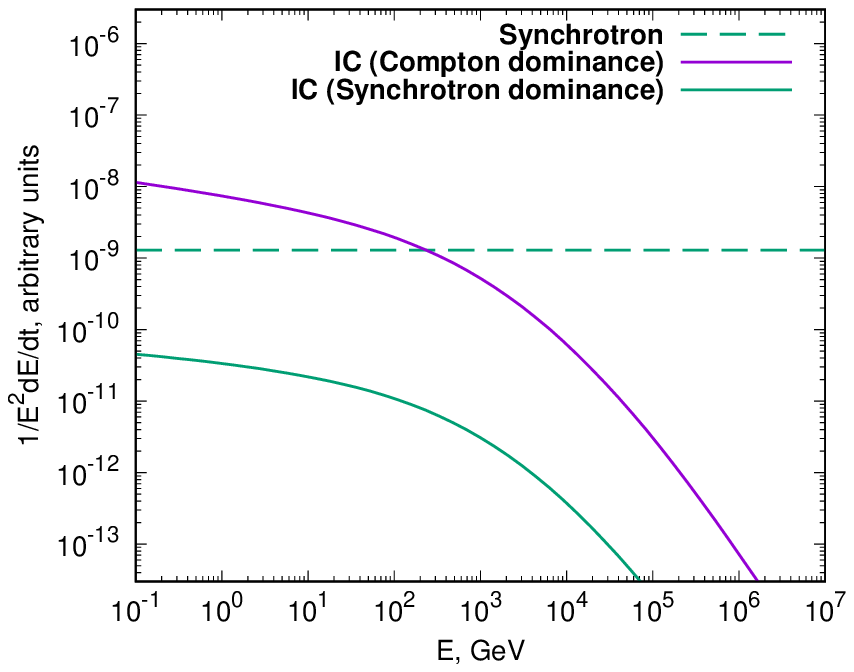}
  \caption{Synchrotron (dashed line) and IC energy losses (solid lines) for two sizes of the production regions, \(R=2.9\times10^{16}\rm\,cm\) (dominant Compton cooling, shown with solid purple line) and \(R=9\times10^{17}\rm\,cm\) (dominant synchrotron cooling, shown with solid green line).}
  \label{fig:klein_nishina_losses}
\end{figure}

Here we have adopted a spherical geometry for the source, and have accounted for enhancement of the density of synchrotron photon in spherical homogeneous sources \cite{1996MNRAS.278..525A}. Two particle spectra were obtained, for \(R=9\times10^{17}\rm cm\) and \(R=2.9\times10^{16}\rm\,cm\). In the former case the synchrotron losses dominate for the assumed \(B=1\rm\,G\). In the case of smaller production region, Compton cooling dominates (see in Fig.~\ref{fig:klein_nishina_losses}). 

\begin{figure}[ht]
  \centering
  \includegraphics[width=2.75in]{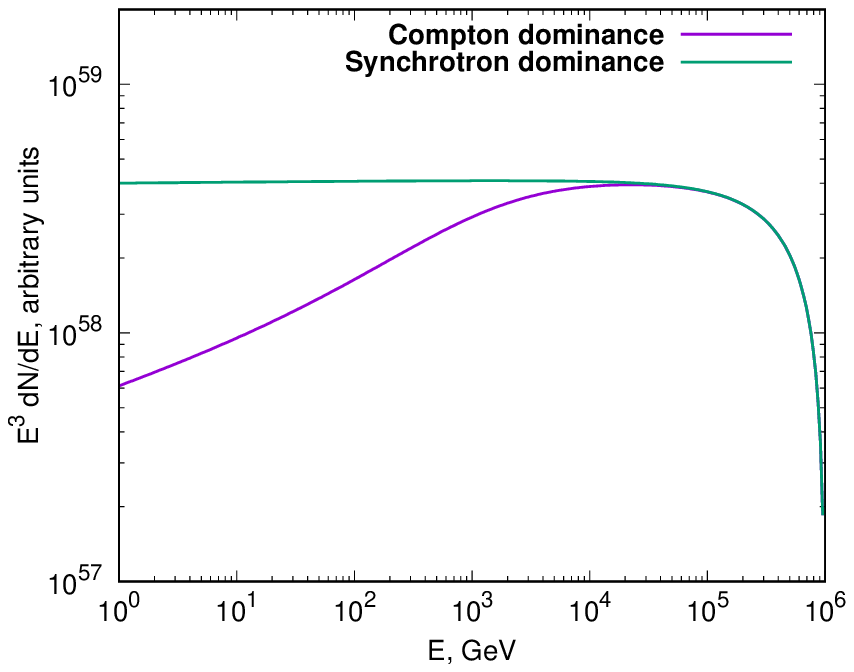}
  \caption{Electron spectrum distributions formed under the dominant synchrotron (shown with solid green line) and IC losses (shown with solid purple line).}
  \label{fig:klein_nishina_particles}
\end{figure}

\begin{figure}[ht]
  \centering
  \includegraphics[width=2.75in]{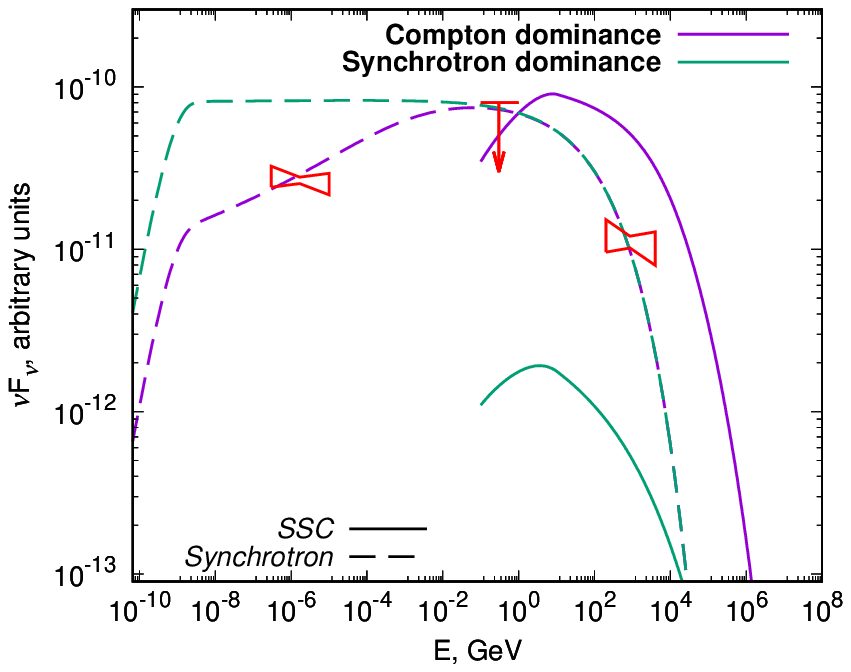}
  \caption{Spectral energy distribution computed for a model that accounts the feedback of Klein-Nishina losses on the particle spectrum. Synchrotron (dashed lines) and IC (solid lines) spectra obtained for two different sized of the production regions, \(R=2.9\times10^{16}\rm\,cm\) (dominant Compton cooling, shown with purple lines) and \(R=9\times10^{17}\rm\,cm\) (dominant synchrotron cooling, shown with green lines); magnetic field was fixed, \(B=1\rm\,G\). Red markers indicate the X-ray and TeV spectra measured during the first night of \grb and relevant \lat upper limits.}
  \label{fig:klein_nishina}
\end{figure}

If the Compton dominance \(Y\) is small, \(R \gg \sqrt{18 L_0 / (B^2 c)}\), then the losses are dominated by the synchrotron and the Klein-Nishina effects results in a steepening of the gamma-ray spectrum (see in Fig.~\ref{fig:klein_nishina}). In the case of high \(Y\), the both the dominant losses and gamma-ray production mechanism  are affected by the Klein-Nishina effect, resulting in a considerable transformation of the particle spectrum (see in Fig.~\ref{fig:klein_nishina_particles}). The particle spectrum transformation is imprinted in the  synchrotron and IC spectra. In Fig.~\ref{fig:klein_nishina} we show the corresponding SED, in which we assumed a bulk Lorentz factor of the production region of \(\Gamma=10\).

\begin{figure}[ht]
  \centering
  \includegraphics[width=2.75in]{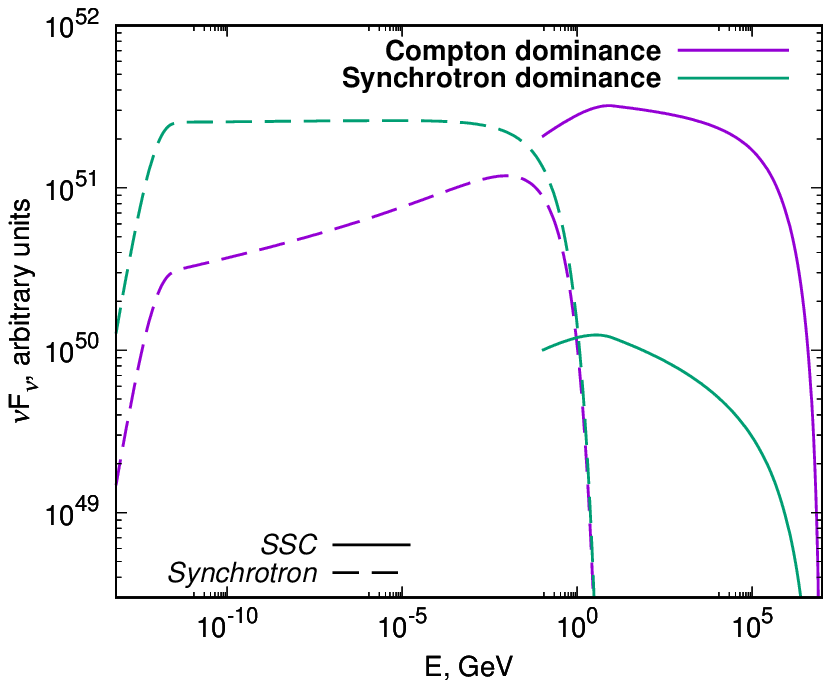}
  \caption{The same as in Fig.~\ref{fig:klein_nishina} for \(B=1\rm\,mG\) (sizes of the production region change to keep the same ratio of the energy densities of target fields).}
  \label{fig:klein_nishina2}
\end{figure}

For the considered model parameters, the spectrum is formed by losses in the transition range, between the Thomson and Klein-Nishina regimes, which resulted in a smaller spectral transformation than predicted by the simple analytic estimates above. However, if particles cool via the Klein-Nishina losses then (i) the SED is dominated by the IC component, (ii) the synchrotron spectrum is hard. We also show in Fig.\ref{fig:klein_nishina} the spectra obtained from \grb in the X-ray and TeV bands during the first night of observations together with \lat upper limits.   We also compute the SED distribution for the case of a weak magnetic field \(B=10^{-3}\rm G\) (shown in Fig.~\ref{fig:klein_nishina2}). It can be seen that the overall SED is not changed considerably.

\subsubsection*{Off-axis explosion}\label{offaxis}
\begin{figure}[ht]
    \centering
    \includegraphics[scale=0.5]{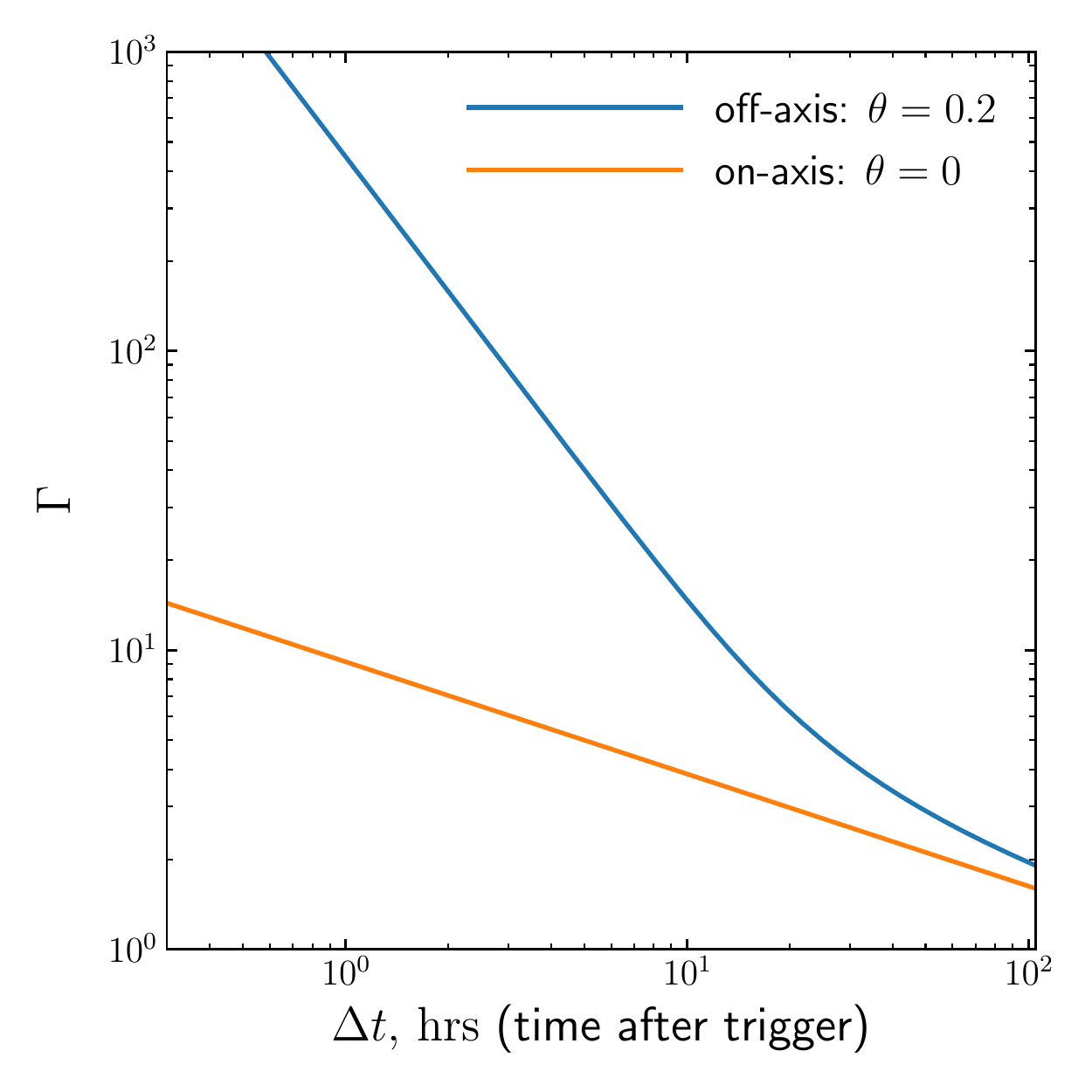}
    \caption{Forward shock Lorentz factors as a function of the detection time for ``off-axis'' and ``on-axis'' cases.}
    \label{fig:offaxis}
\end{figure}

If the observer is located outside the beaming cone of the GRB jet then the consideration of the process changes significantly. As a detailed study of this possibility is beyond the scope of this paper, we only outline here some important features. We assume that the dynamics of the forward shock is well described by the one-dimensional model for a spherical blast wave \cite{1976PhFl...19.1130B}, and that the jet opening angle is small compared to the angle, \(\theta\), between the jet axis and the line-of-sight. In this case, the delay between the trigger and the emission detection is
\be\label{eq:offaxis}
\Delta t \approx \int\limits_0^t {\dif{\tilde{t}}}{\left(1-\beta(\tilde t)\cos\theta\right)} \,,
\ee
where \(t\) is the time elapsed after explosion in the progenitor frame and \(\beta=\sqrt{1-\Gamma^{-2}}\). Comparison of Eqs.~\eqref{eq:trigger} and \eqref{eq:offaxis} shows that the jet-observer angle, \(\theta\), can have a strong influence on the estimate of the bulk Lorentz factor. In Figure~\ref{fig:offaxis} we show calculations corresponding to our baseline model (\(E\iso=2\times10^{50}\rm\,erg\), ISM medium with constant density \(\rho_0=m_p \rm\,cm^{-3}\)) for two orientations of the jet, \(\theta=0\) (``on-axis'') and \(\theta=0.2\rm\,rad\) (``off-axis''). During the fist nigh \(\Delta t =4.3\), the bulk Lorentz factor depends strongly on the angle \(\theta\), for the second and third nights (\(\Delta t=27.2\rm\,hrs\) and \(\Delta t=51.2\rm\,hrs\), respectively), the difference between the ``off-axis'' and ``on-axis'' cases is small.

\end{appendix}

\end{document}